\documentclass[10pt,journal,compsoc]{IEEEtran}

\ifCLASSOPTIONcompsoc
  \usepackage[nocompress]{cite}
\else
  \usepackage{cite}
\fi

\usepackage{url}
\newcommand{\cev}[1]{\reflectbox{\ensuremath{\vec{\reflectbox{\ensuremath{#1}}}}}}

\usepackage{graphicx, subfig}

\usepackage{amsmath}
\usepackage{amsfonts}
\usepackage{multirow}
\usepackage{booktabs}

\begin{document}
\title{DeepMnemonic: Password Mnemonic Genera-\\tion via Deep Attentive Encoder-Decoder Model}


%

\author{Yao Cheng, Chang Xu, Zhen Hai*, and Yingjiu Li
\IEEEcompsocitemizethanks{
\IEEEcompsocthanksitem Yao Cheng is with Huawei International, Singapore.
E-mail: chengyao101@huawei.com
\IEEEcompsocthanksitem Chang Xu is with Data61, CSIRO, Australia.
E-mail: chang.xu@data61.csiro.au
\IEEEcompsocthanksitem *Corresponding author. Zhen Hai is with Institute for Infocomm Research, A*STAR, Singapore.
E-mail: haiz0001@e.ntu.edu.sg
\IEEEcompsocthanksitem Yingjiu Li is with Department of Computer and Information Science, University of Oregon, Eugene, Oregon, United States.
E-mail: yingjiul@uoregon.edu
}
\thanks{Manuscript submitted on June 3, 2019; accepted on April 2, 2020.}}


\IEEEtitleabstractindextext{%
\begin{abstract}
Strong passwords are fundamental to the security of password-based user authentication systems. 
In recent years, much effort has been made to evaluate the password strength or to generate strong passwords.
Unfortunately, the usability or memorability of the strong passwords has been largely neglected.
In this paper, we aim to bridge the gap between strong password generation and the usability of strong passwords.
We propose to automatically generate textual password mnemonics, i.e., natural language sentences, which are intended to help users better memorize passwords. 
We introduce \textit{DeepMnemonic}, a deep attentive encoder-decoder framework which takes a password as input and then automatically generates a mnemonic sentence for the password.
We conduct extensive experiments to evaluate DeepMnemonic on the real-world data sets.
The experimental results demonstrate that DeepMnemonic outperforms a well-known baseline for generating semantically meaningful mnemonic sentences.
Moreover, the user study further validates that the generated mnemonic sentences by DeepMnemonic are useful in helping users memorize strong passwords.
\end{abstract}

\begin{IEEEkeywords}
Password mnemonic generation, deep neural network, attention mechanism, neural machine translation.
\end{IEEEkeywords}}

\maketitle
\IEEEdisplaynontitleabstractindextext

\IEEEraisesectionheading{\section{Introduction}\label{sec:introduction}}

\IEEEPARstart{N}{owadays}, user authentication is the key to ensuring the security of user accounts for most online services, such as social media, e-commerce, and online banking.
Although various authentication schemes have emerged in recent years, e.g., pattern-based or biometric-based authentication, the password-based authentication remains a prevailing choice in most real-world applications, whose security relies on the difficulty in cracking passwords.
Choosing strong passwords becomes extremely important and necessary. 


In reality, service providers often present password policies to aid users in creating strong passwords.
Such policies may require that \textit{a password be longer than a pre-defined minimum length, or contain multiple types of characters (letters, numbers, and special characters)}.
Such policies are expected to guide the generation of passwords that are resistant to password attacks~\cite{proctor2002improving}, but users tend to choose passwords that are easy to memorize in practice~\cite{ur2015added}. 
As a result, password policies may not be as effective as expected~\cite{weir2010testing}.

Much effort has been made to evaluate the strength of passwords~\cite{ma2014study}\cite{melicher2016fast}\cite{wang2016fuzzypsm} or to generate strong passwords~\cite{forget2008improving}\cite{forget2008persuasion}\cite{houshmand2012building}.
Various methods, on the other hand, have been presented to crack passwords, assess whether passwords are sufficiently strong or not, or reduce password guessability~\cite{weir2009password}\cite{veras2014semantic}\cite{castelluccia2012adaptive}. 
However, no rigorous effort has been made to address the \textit{memorability} of strong passwords.
Strong passwords are usually difficult for users to memorize because their entropy values are beyond many users' memorability~\cite{yan2004password}\cite{anderson2014atomic}\cite{miller1956magical}.
The memorability of strong password has become one of the biggest hindrances to the wide adoption of strong passwords in real-world applications.

One possible approach to addressing this problem is to generate passwords that are not only secure but also easy to remember.
Different strategies have been applied to generating word-based memorable passwords, such as pronounceable passwords~\cite{ppg}\cite{mpg}, meaningful passwords~\cite{npg}, and passwords concatenating random words~\cite{npg2}.
Unfortunately, some of such strategies were demonstrated to be vulnerable to certain attacks~\cite{npg}\cite{yang2016empirical}, and moreover, there is no theoretical guarantee that the generated passwords are indeed strong.
In addition, various user-defined rules were adopted for the generation of expression-based memorable passwords~\cite{Kuo:2006:HSM:1143120.1143129}\cite{yang2016empirical}\cite{kiesel2017large}.
One limitation of this approach is that the strength of the generated passwords largely depends on the \textit{uncertainty} of the phrases chosen by users.
This approach also tends to suffer from the low quality of the generation rules~\cite{kiesel2017large}.

Instead of generating strong and memorable passwords, recent effort has been made to assist users in remembering passwords using external tools, such as helper cards~\cite{topkara2007passwords}, hint images~\cite{fukumitsu2010proposal}\cite{juang2012using} and engaging games~\cite{doolani2016improving}.
In particular, Jeyaraman and Topkara~\cite{jeyaraman2005have} proposed a heuristic method that relies on textual hint headlines to deal with the memorability issue of strong passwords.
Given a password, they proposed to \textit{search} and \textit{find} an existing natural language headline that suggests the password from a given corpus (i.e., Reuter Corpus Volume 1).
If the search is unsuccessful, they would then use an external semantic lexicon named \textit{WordNet}~\cite{wordnet} to replace certain words in an existing headline with their synonyms.
In this way, they could find a headline or a variant headline and use it as a hint for memorizing a given password.
This approach is subject to the following limitations:
(i) It can only handle the passwords composed of alphabetic characters. Manual intervention is needed to tackle digits or special characters, which are non-trivial for the composition of strong passwords in reality.
(ii) Meaningful hint headlines are often missing by simply searching the given corpus while creating variant headlines at the word level by using an external lexicon may result in syntactically incorrect or semantically inconsistent headline sentences.
(iii) It only works for short passwords that consist of 6 or 7 characters due to the limited lengths of the headlines in the corpus.

In this work, given strong passwords, we propose to automatically generate \textit{mnemonics}, i.e., natural language sentences, to bridge the gap between security and memorability of the passwords.
Specifically, we introduce a new mnemonic generation system named \textit{DeepMnemonic}, which is capable of generating a semantically meaningful mnemonic sentence for each given textual password.
The core of DeepMnemonic is an attentive encoder-decoder neural network.
It learns to transform any given user password into a mnemonic sentence by first encoding the password character sequence and then decoding the encoded information into the mnemonic sentence via a specially designed attention strategy.
Both encoder and decoder are implemented with recurrent neural networks that can capture the contextual dependency among pair-wise input passwords and mnemonic sentences.
The key insight of DeepMnemonic is inspired by a cognitive psychology theory~\cite{bower1970analysis}, which states that the human ability to memorize and recall certain information is positively influenced by associating additional semantic content with the information~\cite{jeyaraman2005have}.
The natural language mnemonic sentences generated by DeepMnemonic can provide such semantically meaningful content for users to remember and recall the given strong passwords.

Different from the existing textual password hint systems~\cite{jeyaraman2005have}, DeepMnemonic enjoys several unique properties: 
(i) Feature-free: DeepMnemonic is an end-to-end solution, which aims to directly map a given textual password to its corresponding mnemonic sentence, and no manual feature engineering is needed in the learning process. 
(ii) Long-password-friendly: The attention-based recurrent neural network component in DeepMnemonic is capable of identifying salient information in long passwords for generating semantically meaningful sentences. 
(iii) Learning-adaptive: The learning of DeepMnemonic is fitted to different types of password-sentence training pairs so as to process diversified passwords and meet various mnemonic generation requirements.

In summary, we have made the following main contributions in this work:
\begin{itemize}
\item We introduce DeepMnemonic, a mnemonic sentence generation system, to help users remember strong passwords. 
DeepMnemonic utilizes an encoder-decoder neural network model to generate meaningful mnemonic sentences for given strong passwords.
\item We quantitatively evaluate the capability of DeepMnemonic in mnemonic sentence generation. 
Our experimental results show that DeepMnemonic achieves 99.09\% MP (Mnemonic Proportion) and 16.47 BLEU-4 (BiLingual Evaluation Understudy), outperforming an \textit{n-gram} language model baseline (83.62\% MP and 5.09 BLEU-4).  
\item We conduct a user study to qualitatively evaluate the helpfulness of DeepMnemonic. 
The results demonstrate that DeepMnemonic helps users with 54.47\% decrease in time spent on remembering passwords and 57.14\% decrease in recall error measured by the edit distance between each pair of the given password and its recalled version.
\end{itemize}

The rest of this paper is organized as follows. 
Section~\ref{sec:Problem Statement} describes the password mnemonic generation problem and its application scenario.
Section~\ref{sec:Methodology} introduces DeepMnemonic, a deep attentive encoder-decoder system that can generate mnemonic sentences for the given textual passwords.
In Section~\ref{sec:Experiments} and Section~\ref{sec:User Study}, we evaluate DeepMnemonic quantitatively and qualitatively in experiment and user study, respectively.
Section~\ref{sec:Discussion} discusses several usability issues of DeepMnemonic.
Section~\ref{sec:related work} summarizes the related work, and Section~\ref{sec:Conclusion} concludes this paper.

\section{Problem Statement and Application Scenario}
\label{sec:Problem Statement}

\begin{figure*}[hbtp]
\centering
\includegraphics[scale=0.44]{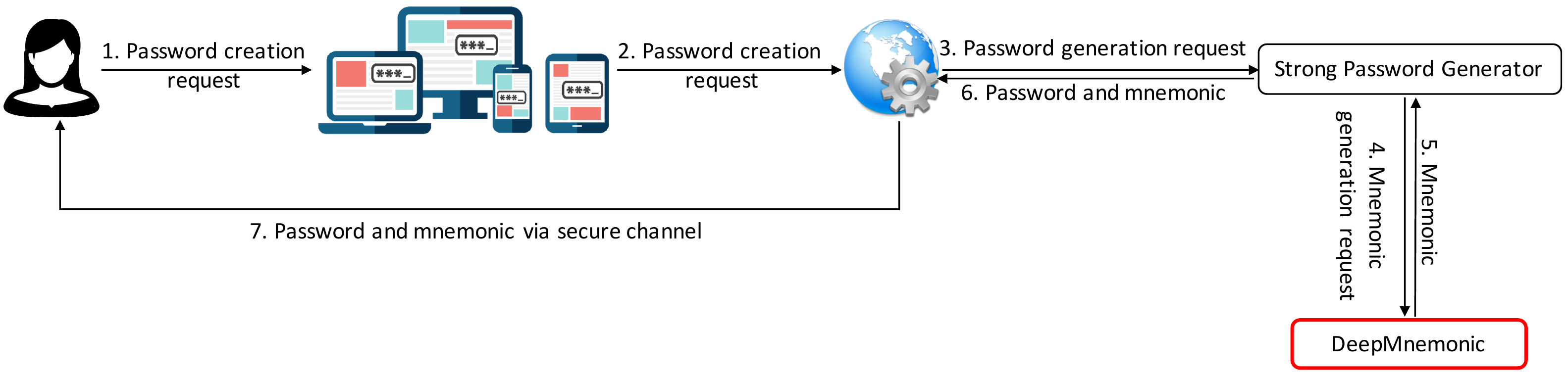}
\caption{An applicable scenario of DeepMnemonic in the password-based authentication system.}
\label{fig:Scenario overview.}
\end{figure*}

Given a strong textual password that often consists of random characters, the aim of this work is to assist common users in remembering the password properly.
According to a famous cognitive psychology theory~\cite{bower1970analysis}, in order to memorize a piece of information, it is helpful to associate some external tools or contextual contents with the information.
More specifically, the cognitive psychology theory~\cite{bower1970analysis} explains that the information to be remembered can be a list of items, for example, a list of characters in a password. 
For the semantic content, it is not limited to a vivid image representing, symbolizing, or suggesting each item of semantic information. 
It indicates that, in order to better memorize a piece of information, users can construct such semantic content and associate it with the information. 
Inspired by the theory, we propose DeepMnemonic, an intelligent system that helps users construct the semantic content by suggesting a meaningful mnemonic sentence corresponding to each character in the password. 
For example, given a text password  \textit{\underline{Tcahm,``Wac?''}} to be remembered, a semantic content can be \textit{\underline{T}he \underline{c}hild \underline{a}sked \underline{h}er \underline{m}other\underline{,`` W}ho \underline{a}re \underline{c}hildren\underline{?''}}, where the first letter of each word in the mnemonic sentence is suggesting and corresponding to each character in the password.

Technically, the problem of password mnemonic generation is closely related to the machine translation problem in natural language processing. 
Both tasks aim to address a sequence-to-sequence learning problem.
Specifically, in machine translation, the goal is to generate a translated sentence in a target language given a sentence in a source language, whereas in our case, we focus on generating a mnemonic sentence for a given password (which is a sequence of characters).
We propose to exploit a neural sequence-to-sequence language model~\cite{bahdanau2014neural} to \textit{translate} a given password (a character sequence) into a meaningful mnemonic sentence (a word sequence).

Formally, let $D$ be a set of $N$ pairs of password and mnemonic sentence $(X_n, Y_n)$ ($n\in\{1,\cdots,N\}$), where $X_n$ denotes a password of $T$ characters, $X_n = x_1, x_2, \cdots, x_T$, where the character $x_t$ ($t\in\{1,\cdots,T\}$) can be alphabetic, punctuation, digital, and special characters, while $Y_n$ refers to a mnemonic sentence of $T$ dictionary terms, $Y_n = y_1, y_2, \cdots, y_T$, where the term $y_t$ ($t\in\{1,\cdots,T\}$) can be words, punctuation marks, numbers, and special tokens.
In the following, we use \textit{words} or \textit{tokens} to refer to dictionary terms.
For each pair of password and mnemonic sentence, a one-to-one mapping relationship exists between the characters of the password and the words of the mnemonic sentence, where each character of the password may appear at certain position of the corresponding mnemonic word.
For example, given a text password \textit{\underline{O,y,slt.}}, a mnemonic sentence can be \textit{\underline{O}h\underline{,} \underline{y}es\underline{,} \underline{s}omething \underline{l}ike \underline{t}hat\underline{.}}, where each alphabetic character of the password corresponds to the first letter of respective words in the mnemonic sentence.

We formulate the mnemonic sentence generation as a natural language generation problem. 
Given a (strong) password $X_n$, the problem is to automatically generate a semantically coherent and meaningful mnemonic sentence $Y_n$ of the same length as $X_n$.
Our solution, named \textit{DeepMnemonic}, employs a neural attentive encoder-decoder language model~\cite{bahdanau2014neural} to generate textual mnemonic sentences for passwords.
DeepMnemonic builds on a sequence-to-sequence learning framework, which encodes an input password (i.e., a sequence of characters) and automatically generates a mnemonic sentence (i.e., a sequence of words) based on the encoded contextual information. 

Figure~\ref{fig:Scenario overview.} shows an applicable scenario of DeepMnemonic in the conventional password-based authentication system.
When a user requires to create a password for registration (step 1 and step 2), the authentication service requests the password generator to generate a strong password (step 3).
Typically, the generated password may not be easy for the user to remember due to the randomness of the password~\cite{yan2004password}.
To improve the memorability of the strong password, the password generator requests DeepMnemonic to produce a mnemonic sentence corresponding to the generated password (step 4).
Once the password generator receives the response from DeepMnemonic (step 5), it returns the generated password as well as the corresponding mnemonic sentence to the service (step 6). 
Finally, the service distributes the password and mnemonic information to the user via a secure channel (step 7). 

Notice that DeepMnemonic does not modify the strong passwords generated by the password generator, and hence does not compromise the security of the generated passwords, under the assumption that users keep both passwords and associated mnemonic sentences secure. 
The passwords' resilience to existing password attacks, such as brute force attacks, dictionary attacks, and guessing attacks, relies on strong password generation which has been well studied independently of our research.
DeepMnemonic does not generate strong passwords but focuses on the usability of strong passwords.
It can generate a corresponding mnemonic sentence for the password so as to overcome the hindrance to the practical use of strong password generation techniques.

\section{Methodology}
\label{sec:Methodology}

\subsection{Overview}

DeepMnemonic applies a deep attentive encoder-decoder learning model~\cite{bahdanau2014neural} to learn a sequence-to-sequence mapping from an input password (i.e., a character sequence) to an output mnemonic sentence (i.e., a word sequence). 
The encoder of DeepMnemonic captures the underlying contextual \textit{meaning} of a password from its sequence of characters, while the decoder generates a corresponding mnemonic sentence based on the encoded content of password.
Typically, the lengths of the given passwords are variable, and strong passwords may consist of long sequences of random characters. 
DeepMnemonic may lose focus in its process, if it treats all characters of an input password equally.
To tackle this issue, DeepMnemonic exploits an attention mechanism~\cite{bahdanau2014neural} to dynamically determine which parts of an input password are more relevant to generating a semantically meaningful mnemonic sentence.

\begin{figure}[hbtp]
\centering
\includegraphics[scale=0.5]{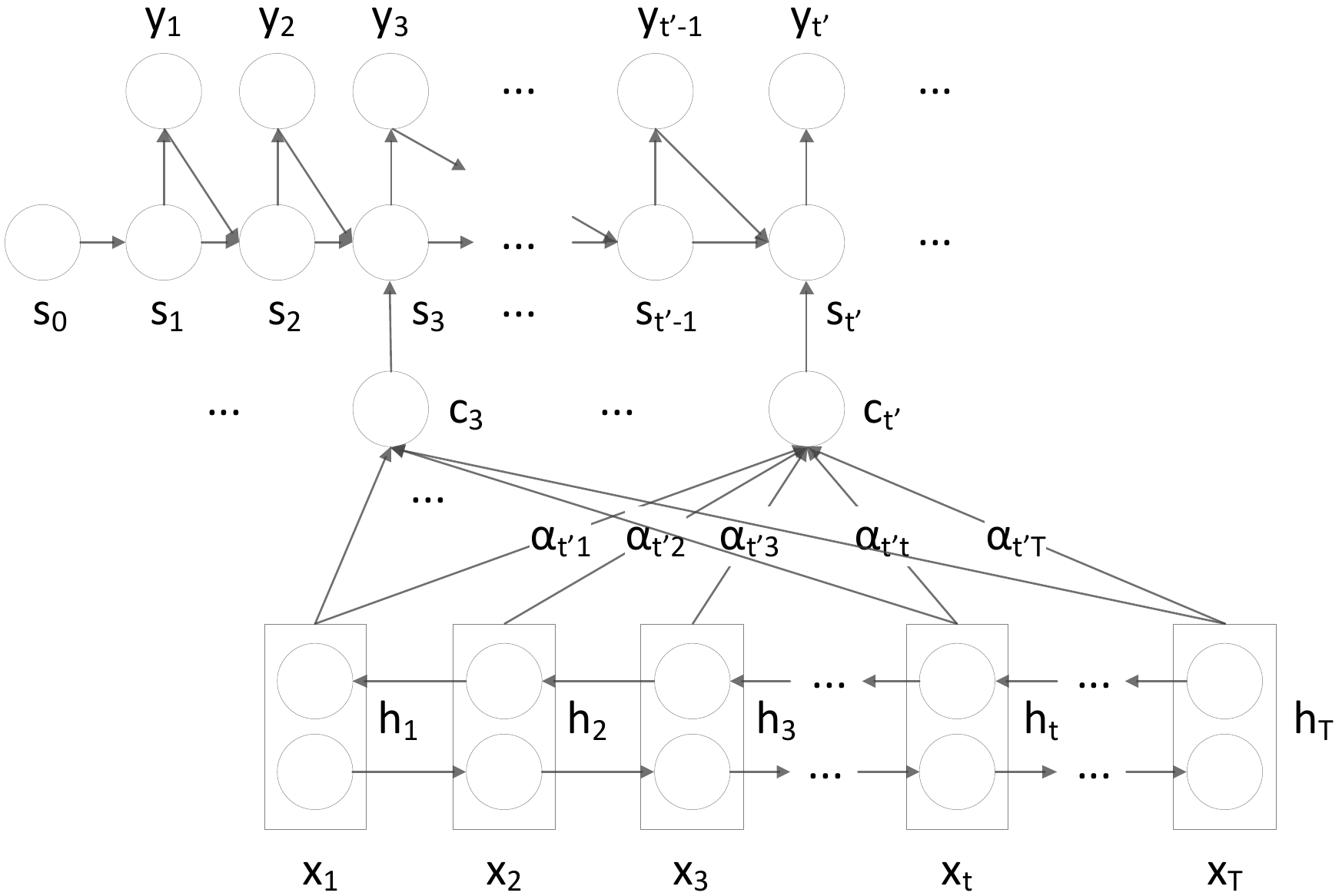}
\caption{The encoder-decoder learning framework of DeepMnemonic.}
\label{deep_mnemonic_model}
\end{figure}

Figure \ref{deep_mnemonic_model} shows the encoder-decoder learning framework of DeepMnemonic. 
The encoder first takes as input a textual password $X_n$ of length $T$,
\begin{equation*}
X_n = (x_1, x_2, \cdots, x_t, \cdots, x_T),
\end{equation*}
where $x_t \in \mathbb{R}^{V_C}, t\in\{1,\cdots,T\}$ denotes a character, and $V_C$ is the size of the input \textit{character vocabulary}.
It derives a hidden context-aware summary or ``meaning'' of the sequence of characters from input password via a bidirectional recurrent neural network (BiRNN)~\cite{schuster1997bidirectional}.
Based on the encoded hidden ``meaning'' of the password, the decoder of DeepMnemonic utilizes an attentive mechanism~\cite{bahdanau2014neural} to generate the corresponding mnemonic sentence $Y_n$ word by word,
\begin{equation*}
Y_n = (y_1, y_2, \cdots, y_t, \cdots, y_T),
\end{equation*}
where $y_t \in \mathbb{R}^{V_W}$ denotes a word, and $V_W$ is the size of the output \textit{word vocabulary}.

\subsection{Encoder}

Given an input password $X_n$, we employ a BiRNN to derive the hidden semantic representation as the meaning of the password. 
BiRNN consists of both forward and backward processes, where the forward process reads the password character sequence in the original order, while the backward process reads the sequence in the reverse order.
BiRNN can thus capture contextual patterns in both directions (left and right) for each password character by summarizing the information not only from the preceding characters but also from the following ones in the sequence.  

Formally, by using the gated recurrent units (GRU)~\cite{cho2014learning}, which is a popular choice for modern RNNs, the forward process of BiRNN computes the hidden state $h_t$ at a given time step $t$ as follows.
First, the forward reset gate $\vec{r}_t$ is computed as,
\begin{equation}
\vec{r}_t = \sigma(\vec{W}^{(r)} E^{(C)} x_t + \vec{U}^{(r)} \vec{h}_{t-1} ),
\end{equation}
where $\sigma$ is the sigmoid function, $E^{(C)} \in \mathbb{R}^{m \times V_C}$ is the character embedding matrix for the password characters, $\vec{W}^{(r)} \in \mathbb{R}^{n \times m}$ and $\vec{U}^{(r)} \in \mathbb{R}^{n \times n} $ are trainable parameter matrices, and $m$ and $n$ refer to the dimensionalities of the character embedding and the hidden state vector, respectively.

Similarly, the forward update gate $\vec{z}_t$ is computed as,
\begin{equation}
\vec{z}_t = \sigma(\vec{W}^{(z)} E^{(C)} x_t + \vec{U}^{(z)} \vec{h}_{t-1} ),
\end{equation}
where $\vec{W}^{(z)} \in \mathbb{R}^{n \times m} $ and $\vec{U}^{(z)} \in \mathbb{R}^{n \times n} $ are trainable parameter matrices.

Then, the forward hidden state vector $\vec{h}_t$ is computed as,
\begin{equation}
\vec{h}_t = (1 - \vec{z}_t ) \circ \vec{h}_{t-1} + \vec{z}_t \circ \vec{\tilde{h} }_t,
\end{equation}
where 
\begin{equation*}
\vec{\tilde{h} }_t = \tanh( \vec{W} E^{(C)} x_t + \vec{U} [\vec{r}_t \circ \vec{h}_{t-1} ] ),
\end{equation*}
both $\vec{W} \in \mathbb{R}^{n \times m} $ and $\vec{U} \in \mathbb{R}^{n \times n} $ are trainable parameters, and $\circ$ refers to an element-wise multiplication.

Following the aforementioned steps, the backward hidden state $\cev{h}_t$ can be computed reversely for any given time step $t$.
Note that the character embedding matrix $E^{(C)}$ is shared for both forward and backward processes of BiRNN.
Next, we concatenate both forward and backward hidden states $\vec{h}_t$ and $\cev{h}_t$ for each time step $t$, and derive a set of overall hidden states $(h_1, \cdots, h_T)$, where 
\begin{equation}
h_t = \left[ \begin{array}{c} \vec{h}_t \\ \cev{h}_t  \end{array} \right].
\end{equation}

\subsection{Decoder}

Based on the encoded hidden states $(h_1, \cdots, h_T)$ of the input password, the decoder of DeepMnemonic employs an attention mechanism to evaluate the importance of individual hidden states.
Then, it derives an overall context-aware representation of the hidden states in order to generate each target word in the corresponding mnemonic sentence.

In particular, for each encoded hidden state $h_t$ at time step $t$, the decoder first computes an attentive weight $\alpha_{t^\prime, t}$ with regard to the contextual state $s_{t^\prime -1}$ at time step $t^\prime$,
\begin{equation}
\alpha_{t^\prime,t} = \frac{\exp(e_{t^\prime, t}) }{ \sum^T_{\tau=1} \exp(e_{t^\prime, \tau}) },
\end{equation}
where 
\begin{equation*}
e_{t^\prime, t} = {v^{(a)}}^T \tanh( W^{(a)} s_{t^\prime -1} + U^{(a)} h_t)
\end{equation*}
is an alignment model that evaluates the relevance between the input hidden state $h_t$ and the previous output hidden state $s_{t^\prime -1}$.
The $v^{(a)} \in \mathbb{R}^{n^\prime}$, $W^{(a)} \in \mathbb{R}^{n^\prime \times n}$, and $U^{(a)} \in \mathbb{R}^{n^\prime \times 2n}$ are trainable parameters, and $n^\prime$ refers to the dimensionality of the hidden state vector of the attention layer.

Then, the attentive context vector $c_t$ is computed via a weighted sum,
\begin{equation}
c_{t^\prime} = \sum^T_{t=1} \alpha_{t^\prime, t} h_t.
\end{equation}

We can then compute the hidden state $s_{t^\prime}$ of the decoder at time $t^\prime$ given the hidden states of encoder via a decoder GRU as below,
\begin{equation}
s_{t^\prime} =  (1-z_{t^\prime}) \circ s_{t^\prime -1} + z_{t^\prime} \circ \tilde{s}_{t^\prime},
\end{equation}
where 
\begin{equation*}
\tilde{s}_{t^\prime} = \tanh( W E^{(W)} y_{t^\prime-1} + U[r_{t^\prime} \circ s_{t^\prime -1}] + C c_{t^\prime} ),
\end{equation*}
\begin{equation*}
z_{t^\prime} = \sigma( W^{(z)} E^{(W)} y_{t^\prime -1} + U^{(z)} s_{t^\prime -1} + C^{(z)} c_{t^\prime} ),
\end{equation*}
\begin{equation}
r_{t^\prime} = \sigma( W^{(r)} E^{(W)} y_{t^\prime -1} + U^{(r)} s_{t^\prime -1} + C^{(r)} c_{t^\prime} ).
\end{equation}
In the above formulas, $E^{(W)} \in \mathbb{R}^{m \times V_W} $ is the word embedding matrix for the output mnemonic sentence;
$W, W^{(z)}, W^{(r)} \in \mathbb{R}^{n \times m}$, 
$U, U^{(z)}, U^{(r)} \in \mathbb{R}^{n \times n}$ and
$C, C^{(z)}, C^{(r)} \in \mathbb{R}^{n \times 2n}$ are all trainable parameters. 
$m$ and $n$ are the dimensionalities of the word embedding and the hidden state vector, respectively.
Note that the initial hidden state of decoder $s_0$ is computed as,
\begin{equation}
s_0 = \tanh(W^{(s)} \cev{h}_1),
\end{equation}
where $W^{(s)} \in \mathbb{R}^{n \times n}$ is the parameter matrix.

Next, given the decoder state $s_{t^\prime}$, the attentive context vector $c_{t^\prime}$, and the previous generated word $y_{t^\prime -1}$,
we follow~\cite{bahdanau2014neural} and define the probability for generating the target word $y_{t^\prime}$ as follows,

\begin{equation}
p(y_{t^\prime} | s_{t^\prime}, c_{t^\prime}, y_{t^\prime -1}) \propto \exp(y^T_{t^\prime -1} W^{(o)} l_{t^\prime}),
\end{equation}
where 
\begin{equation*}
l_{t^\prime} = [\max\{\tilde{l}_{t^\prime, 2k-1}, \tilde{l}_{t^\prime, 2k} \} ]^T_{k=1,\cdots, K},
\end{equation*}
and $W^{(o)} \in \mathbb{R}^{V_W \times K}$.
The hidden state $\tilde{l}_{t^\prime}$ is computed as follows,

\begin{equation}
\tilde{l}_{t^\prime} = U^{(o)} s_{t^\prime} + V^{(o)} E^{(W)} y_{t^\prime -1} + C^{(o)} c_{t^\prime},
\end{equation}
where $U^{(o)} \in \mathbb{R}^{2K \times n}$, $V^{(o)} \in \mathbb{R}^{2K \times m}$, and $C^{(o)} \in \mathbb{R}^{2K \times 2n}$ are trainable parameters.

\section{Experiments}
\label{sec:Experiments}

We conducted quantitative experiments on a large-scale publicly available dataset to evaluate DeepMnemonic for mnemonic sentence generation.
We also carried out a case study to understand the influence of password lengths and digital/special characters on the utility of DeepMnemonic.
In addition, we provided a visualized understanding of the attention mechanism in DeepMnemonic.

\subsection{Dataset}

The ``Webis-Simple-Sentences-17'' is a publicly available data source for analyzing \textit{expression-based passwords}, and the sentences in the dataset are similar to the human-chosen mnemonics in terms of syllable distribution~\cite{kiesel2017large}.
After filtering out those sentences that contain non-English words, we derived a password from each remaining sentence by concatenating the first letter of each word and the rest special characters (including punctuation marks and numerical digits) in the original order as shown in the sentence.
For example, a password \textit{\underline{O,y,slt.}} can be derived from the following sentence, \textit{\underline{O}h\underline{,} \underline{y}es\underline{,} \underline{s}omething \underline{l}ike \underline{t}hat\underline{.}}
In this way, we collected totally 500,000 pairs of passwords and sentences, forming a ground-truth dataset for building DeepMnemonic.
It is worth noting that, in creating the ground truth data, it is also possible to derive a password by concatenating the last (or a middle) letter of each word and the rest special characters in each sentence.

In preprocessing the ground-truth data, all words of each sentence were converted to lowercase, which was helpful to reduce the size of vocabulary for the language generation.
In particular, by following the given passwords, we can easily restore the corresponding words of the generated mnemonic sentences to uppercase wherever applicable.
In addition, extra symbols, such as $<$s$>$ and $<$/s$>$, were inserted at the start and the end of each sentence to indicate its boundary.
A special \textit{unknown} symbol, i.e., $<${\scriptsize UNK}$>$, was included in the vocabulary, and would be used for the generation of sentences when no appropriate words can be predicted.
In the pairwise ground-truth dataset, the minimum and maximum lengths of passwords are 8 and 16, respectively, which are compatible with the password length requirements in many authentication systems.

Among the ground truth dataset, we randomly selected 450,000 pairs as training data, and used the rest 50,000 for testing.
Among the training data, 20\% of the pairs were randomly held out as validation data for model selection.
After preprocessing, we obtained a vocabulary of 109,584 words (tokens) for the mnemonic sentence generation task.

\subsection{Experimental Setting}

In DeepMnemonic, the hidden layer sizes of both encoder and decoder were set to 256, which were optimally selected using the validation set.
The dropout strategy has been previously shown effective to prevent neural network models from overfitting~\cite{srivastava2014dropout} and the dropout rate was set to 0.2 in our experiments.

Note that no extra information, e.g., pre-defined generation rules, is required beyond training data.
Once the training process is done, the DeepMnemonic can be used to automatically generate a mnemonic sentence for any given password. 

\subsubsection{Beam Search}
Simply generating the \textit{best} word that achieves the highest predictive probability at each time step may not always result in an overall semantically meaningful sentence in practice.
Therefore, a left-to-right beam search strategy~\cite{koehn2004pharaoh} was employed to find the most likely mnemonic sentence~\cite{sutskever2014sequence}\cite{bahdanau2014neural} for each given password.

In particular, the beam search based decoder stores a predetermined number $b$ (beam width) of partial sentences, where each partial sentence is a prefix of a candidate mnemonic. 
At each time step, each partial sentence in the beam grows with a possible mnemonic word from the decoder vocabulary. 
Clearly, this process would greatly increase the number of candidate sentences.
To overcome this issue, only $b$ most likely candidates are maintained in terms of their predicted probabilities. 
Once the end of sentence symbol is appended to a candidate mnemonic, it is included in the set of full mnemonic sentences. 
In general, the wider the beam width $b$ is, the more candidate mnemonics the decoder searches for, and thus the better results could be achieved.

\subsubsection{Baseline}

The $n$-gram language model is widely known as one of the dominant methods for probabilistic language modeling~\cite{manning1999foundations}.
As a non-parametric learning method, it primarily utilizes the preceding sequence of $n-1$ words to estimate a conditional probability for the prediction of the current word.
Various values of $n$ were tested for $n$-gram, and the \textit{bigram} language model ($n=2$) achieved decent generation performance and was chosen as the baseline to benchmark the proposed DeepMnemonic.

The bigram language model depends on the preceding word for estimating the conditional probability, and then generates the current word that has the highest probability.
Therefore, the bigram model is unable to automatically figure out the relationship between a given password and its mnemonic sentence, i.e., the correspondence between the alphabetic characters of the password and the first letters of mnemonic words in the sentence.
To properly apply the bigram language model to password mnemonic generation, we manually adopted a pre-defined generation rule.
Specifically, to generate a mnemonic word, it is required that the word not only achieve the highest conditional probability (given its preceding word), but its first letter also must be identical to the corresponding character in the given password.

\subsubsection{Evaluation Metrics}

Two metrics were used to evaluate the quality of generated mnemonic sentences via DeepMnemonic and the baseline method.
One metric is BLEU (BiLingual Evaluation Understudy), which is one of the most popular metrics for evaluating the quality of machine translation task in natural language processing~\cite{papineni2002bleu}.
Following~\cite{bahdanau2014neural}, we used BLEU to evaluate the quality of the generated mnemonic sentences with respect to ground-truth sentences in the test set, where the quality refers to the correspondence between each pair of the generated sentence and the ground-truth sentence.
In other words, the closer the generated mnemonic sentence is to the ground-truth sentence, the more meaningful and consistent it is.
BLEU-$n$ is defined as follows (the higher, the better).

\begin{equation*}
BLEU{\text -}n = B \cdot exp(\frac{1}{N^\prime}\sum^{N^\prime}_{n=1} log p_n),
\end{equation*}
where $N^\prime$ is the maximum length of $n$-grams and $N^\prime=4$ was used in the experiments. 
$B$ refers to brevity penalty, and is computed as
\begin{equation*}
B = \left\{ \begin{array}{rcl}  1 \quad \quad \quad \quad  if \quad c > r \\ e^{1-r/c}  \quad \quad if \quad c \leq r \end{array}\right\},
\end{equation*}
where $c$ is the size of the generated mnemonic set, and $r$ is the size of the ground truth sentence set.
The modified n-gram precision $p_n$ is computed as
\begin{equation*}
p_n = \frac{\sum_{C \in \mathcal{H} } \sum_{\text{n-gram} \in C} Count_{clip}(\text{n-gram}) } {\sum_{C^\prime \in \mathcal{H} } \sum_{\text{n-gram}^\prime \in C^\prime} Count(\text{n-gram}^\prime) },
\end{equation*}
where $\mathcal{H}$ is the set of generated mnemonic sentences, $Count(\text{n-gram}^\prime)$ is the number of $\text{n-gram}^\prime$s in a generated mnemonic, and $Count_{clip}(\text{n-gram})$ is the clipped number of the $\text{n-gram}$ of a generated mnemonic with regard to the corresponding ground-truth mnemonic sentence.

The other metric, Mnemonic Proportion (MP), is specifically defined for testing the matching proportion between pair-wise passwords and mnemonic sentences.
Specifically, given each pair of password $X_n$ and generated mnemonic sentence $Y_n$, MP calculates the proportion of the cases where each word $y_t$ of mnemonic $Y_n$ does start with the corresponding character $x_t$ in password $X_n$, as shown below:

\begin{equation*}
MP = \frac{\sum_{n=1}^{N} MP_n}{N},
\end{equation*}
where $N$ is the size of the test set.
Then $MP_n$ is defined as, 
\begin{equation*}
MP_n = \frac{|\{y_t|\,y_t \in Y_n \, \text{and} \, FirstLetter(y_t)=x_t\}|}{|X_n|}
\end{equation*}
where $|X_n|$ is the length of password $X_n$, and $FirstLetter(\cdot)$ is a function that returns the first letter of a word.

\subsection{Experimental Results}

This section reports the results of mnemonic sentence generation via DeepMnemonic and the baseline model in terms of MP and BLEU.
DeepMnemonic ran on Nvidia Tesla P100 GPU with 16GB memory. 
Its training process with 450,000 training data pairs took around 15 hours, and the inference on the entire test data set with 50,000 passwords took about 5 minutes.

\subsubsection{MP}
\label{MP}

Table~\ref{table: MP} shows the MP results of the DeepMnemonic and bigram language model (the higher, the better).
As we can see, DeepMnemonic achieves much better results compared to Bigram over different beam widths.

\begin{table}[ht]
\centering
\caption{The MP results of DeepMnemonic and bigram language model (Bigram) given beam width ($b$) of 1 and 5.}
\label{table: MP}
\begin{tabular}{ccc}
\toprule
             & DeepMnemonic & Bigram  \\ \midrule
$b=1$ & 99.09\%      & 83.62\% \\ \midrule
$b=5$ & 99.15\%      & 98.44\% \\ \midrule
\end{tabular}
\end{table}

Given beam width $b=1$, DeepMnemonic attains an MP value of 99.09\%, while Bigram only achieves 83.62\% MP. 
When the beam width increases to 5, the Bigram MP increases up to 98.44\%, but is still lower than DeepMnemonic.
Surprisingly, increasing the beam width does not lead to significant gain to DeepMnemonic.
This suggests that DeepMnemonic can achieve a high MP value given a small beam width  $b=1$, and it is not sensitive to the choice of width of the beam search.

\subsubsection{BLEU}

This section reports the BLEU-$n$ scores for the generated sentences with regard to ground-truth mnemonic sentences~\cite{papineni2002bleu}\cite{sutskever2014sequence}.
Table~\ref{table: BLEU scores} lists the BLEU-$n$ results with the order $n\in\{1,2,3,4\}$ ($N^\prime=4$) for DeepMnemonic and Bigram given different beam width $b\in\{1,5\}$.
Overall, DeepMnemonic outperforms Bigram significantly and consistently in all cases.

\begin{table}[ht]
\centering
\caption{The BLEU scores of DeepMnemonic and bigram language model (Bigram) given beam width ($b$) 1 and 5. }
\label{table: BLEU scores}
\begin{tabular}{ccccc}
\toprule
\multirow{2}{*}{} & \multicolumn{2}{c}{DeepMnemonic} & \multicolumn{2}{c}{Bigram} \\  
                  & $b=1$          & $b=5$          & $b=1$       & $b=5$       \\ \midrule
BLEU-1            & 45.17           & 45.29           & 28.82        & 37.22        \\ \midrule
BLEU-2            & 30.26           & 30.38           & 14.72        & 22.07        \\ \midrule
BLEU-3            & 22.33           & 22.44           & 8.49         & 14.29        \\ \midrule
BLEU-4            & 16.47           & 16.57           & 5.09         & 9.48         \\ \midrule
\end{tabular}
\end{table}

If beam width is fixed at $b=1$ or $b=5$ , the BLEU-$n$ score for each model decreases as the order $n$ grows from 1 to 4.
This is expected, as increasing the order $n$  results in a stricter evaluation of BLEU for the generated sentences, which is consistent with the findings in~\cite{papineni2002bleu}.
In other words, not only are the words between each pair of the generated and the ground-truth sentences required to be identical, but the order of the words in the $n$-gram also needs to be exactly the same.

DeepMnemonic achieves the best BLEU-1 of 45.29 compared to the baseline Bigram given the beam width $b=5$.
Roughly, this means that about 45 out of 100 generated mnemonic words (unigrams) are identically matched to the ground truth.
In addition, the results again show that DeepMnemonic is robust to the width of the beam search.  
In contrast, the BLEU-$n$ values (from 1 to 4) of Bigram improve significantly by increasing the beam width from  $b=1$ to $b=5$. 
But the best BLEU-1 (37.22) of Bigram at $b=5$ is still lower than DeepMnemonic at  $b=1$.

In addition, it is worth noting that BLEU-$n$ is designed to measure the \textit{correspondence} between each pair of the generated mnemonic sentence and ground-truth sentence.
However, it is possible that a generated mnemonic sentence is helpful to assist users in memorizing a given password, but it may have a low BLEU-$n$ score if the generated sentence does not match to its ground truth very well.
To mitigate this issue, Section~\ref{sec:User Study} further conducts a user study about the helpfulness of DeepMnemonic, where participants are invited to memorize and recall the assigned passwords using mnemonic sentences generated by DeepMnemonic.

\subsection{Case Study}

In this section, we conduct a case study to provide a complementary understanding of the generated mnemonic sentences.
The case study reveals the influence of password lengths and digital/special characters on mnemonic generation.
Moreover, we visualize the effectiveness  of the attention mechanism in DeepMnemonic to explain its semantic meaningfulness.

Given a list of 10 randomly selected passwords of different lengths, Table~\ref{table: Mnemonic Generation Examples} shows the corresponding mnemonic sentences generated via DeepMnemonic and Bigram methods.
Note that the row ``Original'' shows the ground-truth sentences from our ground-truth dataset.

\begin{table*}[]
\centering
\caption{Randomly selected passwords and the corresponding mnemonic sentences generated by DeepMnemonic and Bigram (\textit{Original} means that the mnemonic sentences come from the ground-truth dataset).}
\label{table: Mnemonic Generation Examples}
\begin{tabular}{ccll}
\toprule
No.                      & Length              & Items        & Values                                                                                                                                                                                                                                                                                      \\ \midrule
\multirow{4}{*}{Case 1}                  & \multirow{4}{*}{Short}  & Password     & Y m s , b t o .                                                                                                                                                                                                                                                                             \\
                         &                     & Original     & you'll miss some , but that's okay .                                                                                                                                                                                                                                                        \\
                         &                     & DeepMnemonic & you may say , but that's ok .                                                                                                                                                                                                                                                               \\
                         &                     & Bigram       & you might say , but the other .                                                                                                                                                                                                                                                             \\ \midrule
\multirow{4}{*}{Case 2}  & \multirow{4}{*}{Short}  & Password     & S I b a a / e ?                                                                                                                                                                                                                                                                             \\
                         &                     & Original     & should i buy a acoustic / electric ?                                                                                                                                                                                                                                                        \\
                         &                     & DeepMnemonic & should i be an author / editor ?                                                                                                                                                                                                                                                            \\
                         &                     & Bigram       & so i bought at a / etc ?                                                                                                                                                                                                                                                                    \\ \midrule
\multirow{4}{*}{Case 3}  & \multirow{4}{*}{Short}  & Password     & T f i d i F 1 .                                                                                                                                                                                                                                                                             \\
                         &                     & Original     & this finding is diagrammed in figure 1 .                                                                                                                                                                                                                                                    \\
                         &                     & DeepMnemonic & the festival is due in february 1937 .                                                                                                                                                                                                                                                      \\
                         &                     & Bigram       & the first i do it for 1 .                                                                                                                                                                                                                                                                   \\ \midrule
\multirow{4}{*}{Case 4}  & \multirow{4}{*}{Medium} & Password     & B I j c d t * g * t .                                                                                                                                                                                                                                                                       \\
                         &                     & Original     & but i just can't do the * giddy * thing .                                                                                                                                                                                                                                                   \\
                         &                     & DeepMnemonic & but i just can't do the * good * thing .                                                                                                                                                                                                                                                    \\
                         &                     & Bigram       & but i just can't do they * go $<${\tiny UNK}$>$ $<${\tiny UNK}$>$ $<${\tiny UNK}$>$                                                                                                                                                                                  \\ \midrule
\multirow{4}{*}{Case 5}  & \multirow{4}{*}{Medium} & Password     & T c a h m , `` W a c ? '' \\
                         &                     & Original     & the child asks his mother , `` what are circles ? ''                                                                                                                                                                                                                                          \\
                         &                     & DeepMnemonic & the child asked her mother , `` who are children ? '' \\
                         &                     & Bigram       & they can also have more , `` we are called ? ''                                                                                                                                                                                                                                               \\ \midrule
\multirow{4}{*}{Case 6}  & \multirow{4}{*}{Medium} & Password     & D t m t b h s 7 c i 2 h ?                                                                                                                                                                                                                                                                   \\
                         &                     & Original     & does this mean the book has sold 7 copies in 24 hours ?                                                                                                                                                                                                                                     \\
                         &                     & DeepMnemonic & does this mean that because he sees 75 casualties in 24 hours ?                                                                                                                                                                                                                             \\
                         &                     & Bigram       & during the more than before he said 70 countries in 2006 have ?                                                                                                                                                                                                                             \\ \midrule
\multirow{4}{*}{Case 7}  & \multirow{4}{*}{Medium} & Password     & B Z e p a l i h h t c h .                                                                                                                                                                                                                                                                   \\
                         &                     & Original     & but zeus ever pursued and longed in his heart to catch her .                                                                                                                                                                                                                                \\
                         &                     & DeepMnemonic & but zelotes ever pushed a line in his head to crucify him .                                                                                                                                                                                                                                 \\
                         &                     & Bigram       & but z2 $<${\tiny UNK}$>$ $<${\tiny UNK}$>$ $<${\tiny UNK}$>$ $<${\tiny UNK}$>$ $<${\tiny UNK}$>$ $<${\tiny UNK}$>$ $<${\tiny UNK}$>$ $<${\tiny UNK}$>$ $<${\tiny UNK}$>$ $<${\tiny UNK}$>$ $<${\tiny UNK}$>$ \\ \midrule
\multirow{4}{*}{Case 8}  & \multirow{4}{*}{Medium} & Password     & T , I w l t r t l o K J .                                                                                                                                                                                                                                                                   \\
                         &                     & Original     & today , i would like to recognize the life of ken jablonski .                                                                                                                                                                                                                               \\
                         &                     & DeepMnemonic & today , i would like to read the letter of karl james .                                                                                                                                                                                                                                     \\
                         &                     & Bigram       & this , it was like to read the lord of knowledge $<${\tiny UNK}$>$ $<${\tiny UNK}$>$                                                                                                                                                                                        \\ \midrule
\multirow{4}{*}{Case 9}  & \multirow{4}{*}{Long} & Password     & D t s w i N - D , l a t a p .                                                                                                                                                                                                                                                               \\
                         &                     & Original     & during the six weekends in new - delhi , lunch and tea are provided .                                                                                                                                                                                                                       \\
                         &                     & DeepMnemonic & during the second week in november - december , lakes awoke to a pond .                                                                                                                                                                                                                     \\
                         &                     & Bigram       & during the same way is no - day , look at the air pollution .                                                                                                                                                                                                                               \\ \midrule
\multirow{4}{*}{Case 10} & \multirow{4}{*}{Long} & Password     & A : Y , b t e n t b a i a l .                                                                                                                                                                                                                                                               \\
                         &                     & Original     & ac : yes , but the egg needs to be altered in a lab.  \\
                         &                     & DeepMnemonic & alan : yeah , but that's exactly not the best album in a league .                                                                                                                                                                                                                           \\
                         &                     & Bigram       & a : yes , but the early next to be an important as long .   
\\ \midrule
\end{tabular}
\end{table*}

\subsubsection{Password Length}

It can be observed that, for short passwords, all the sentences generated by either DeepMnemonic or Bigram match all the password characters (i.e., 100\%  MP).
However, as the password length grows, the unknown token ``$<${\scriptsize UNK}$>$'' begins to appear more frequently in the mnemonic sentences generated by Bigram compared to DeepMnemonic.
The main reason is that, when generating a sentence, it is sometimes difficult for the Bigram model to find a word that not only identically matches the given password character but also has the conditional probability greater than zero given the preceding word.
For example, Bigram begins to generate the unknown token ``$<${\scriptsize UNK}$>$'' from the third position onwards in Case 7.
Given the first generated word ``but'', the model identifies the next word ``z2'', which has the highest probability and also starts with the character ``z'' of the given password.
However, when continuing to generate the next word based on ``z2'', Bigram fails to find any words that start with the password character ``e'' and also have the positive conditional probability. 
As a result, Bigram generates an unknown token instead. 
In contrast, DeepMnemonic does not have such issue, and for each given password, it can complete the automatic generation of an entire meaningful sentence.

\begin{figure}[h]
\centering
\includegraphics[scale=0.54]{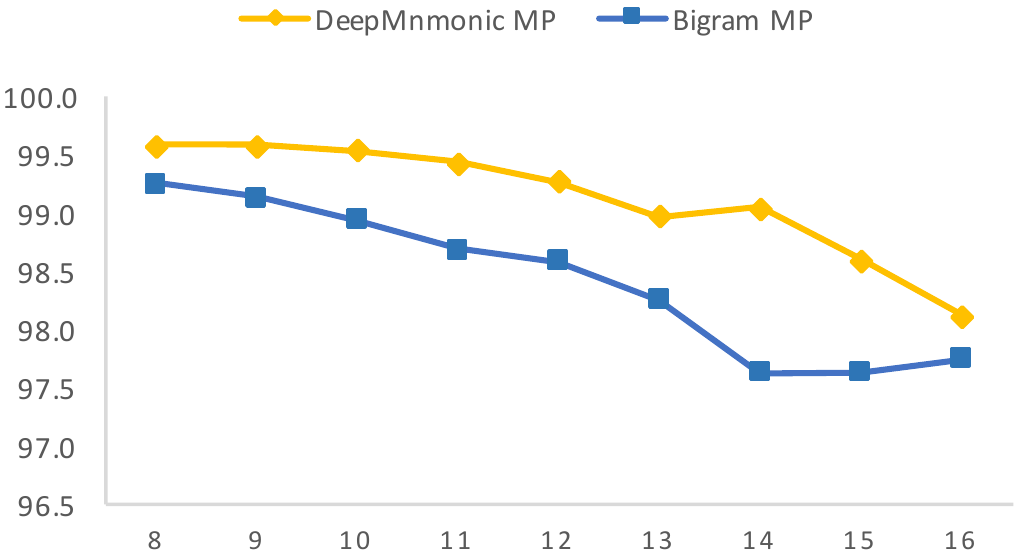}
\caption{Impact of the password length on MP values ($\%$) given $b=5$.}
\label{fig: Impact of password length.}
\end{figure}

Figure~\ref{fig: Impact of password length.} plots the impact of password lengths on MP values at $b=5$. 
We can observe that for both Bigram and DeepMnemonic, MP values decrease as the passwords become longer.
The longer the passwords are, the more difficult the task of generating semantically sensible sentences is.
It is clear that the MP values of DeepMnemonic are always better than those of Bigram at different password lengths, suggesting that DeepMnemonic generates better mnemonic sentences from the given passwords.
For example, the mnemonic sentence generated by DeepMnemonic in Case 9 in Table~\ref{table: Mnemonic Generation Examples} is more sensible and memorable than that by Bigram, although both sentences match all characters of the given password.

\subsubsection{Digital and Special Characters}

In many password generation systems, digital and special characters play an important role in creating strong passwords.
DeepMnemonic is able to handle such characters when generating mnemonic sentences for strong passwords.

As shown in Case 3 of Table~\ref{table: Mnemonic Generation Examples}, DeepMnemonic generates a digital sequence ``1937'' from number ``1'' in the given password, and then it concatenates previously generated words to form a meaningful expression, ``in february 1937'', for the given password subsequence ``iF1''.
In contrast, Bigram generates the words ``it for 1'' for the same subsequence, which is less readable and not much helpful for remembering the password. 
Overall the mnemonic sentence generated by Bigram is not as meaningful as the sentence by DeepMnemonic.
In Case 6, DeepMnemonic generates a sensible mnemonic phrase ``75 causalities in 24 hours'' from a password segment ``7ci2h'', and also produces an overall semantically meaningful mnemonic sentence. 
Although Bigram generates a meaningful phrase ``70 countries in 2006 have'' from the same password segment, unfortunately, it ends up with a grammatically incorrect and semantically inconsistent sentence.

In general, it is challenging to handle special characters of passwords such as ``/'' and ``*'' during generating mnemonic sentences.
For example, in Case 2, given a slash character ``/'' between ``a'' and ``e'' of the password, DeepMnemonic generates a sensible phrase `` author / editor'', where ``/'' normally represents ``or'', while the Bigram model outputs a strange phrase ``a / etc'' for the same password segment.
In Case 4 of Table~\ref{table: Mnemonic Generation Examples}, following the password segment ``*g*'', DeepMnemonic generates a reasonable phrase ``*good*''.
Though it is not identical to the original one (``*giddy*''), semantically, this phrase can be used to highlight the meaning of ``good'' in the generated sentence. 
However, Bigram fails to generate either a short sensible phrase or an entire semantically meaningful sentence.

One more interesting example is the colon symbol ``:'' in Case 10, which is typically used to indicate the start of an utterance.
Surprisingly, DeepMnemonic generates a name ``alan'' (Alan) for the character ``A'' in front of the colon ``:'', and then generates an utterance for the subsequence of characters following the colon symbol.
Bigram fails again to handle this special character.

\subsubsection{Attentive Generation of Mnemonic Sentences}

Generally, a mnemonic sentence, which not only literally covers the characters of the given password, but is also semantically meaningful, is more useful for users to memorize every password character properly.

As shown in Case 8 of Table~\ref{table: Mnemonic Generation Examples}, DeepMnemonic generates a mnemonic sentence, ``today, i would like to read the letter of karl james.'' from password ``T,IwltrtloKJ.''
Although this is not identical to the ground-truth sentence, i.e., ``today, i would like to recognize the life of ken jablonski.'', our generated mnemonic sentence seems easier for users to follow and recall the corresponding password.
To be specific, DeepMnemonic decodes the given password segment ``KJ'' as a name, ``karl james'', and surprisingly, it turns out to be a new person name that does not even appear in the training data.
This demonstrates the ability of DeepMnemonic to capture the semantic context of input passwords as well as the relationship between each pair of password and mnemonic sentence.
This also shows that the attention mechanism in DeepMnemonic captures not only the full view of an input password context but also its salient parts, both of which are useful for mnemonic word inference at each generation step.

\begin{figure}
\subfloat[Case 5.]{\label{fig: Case 5.}
\includegraphics[width=0.23\textwidth]{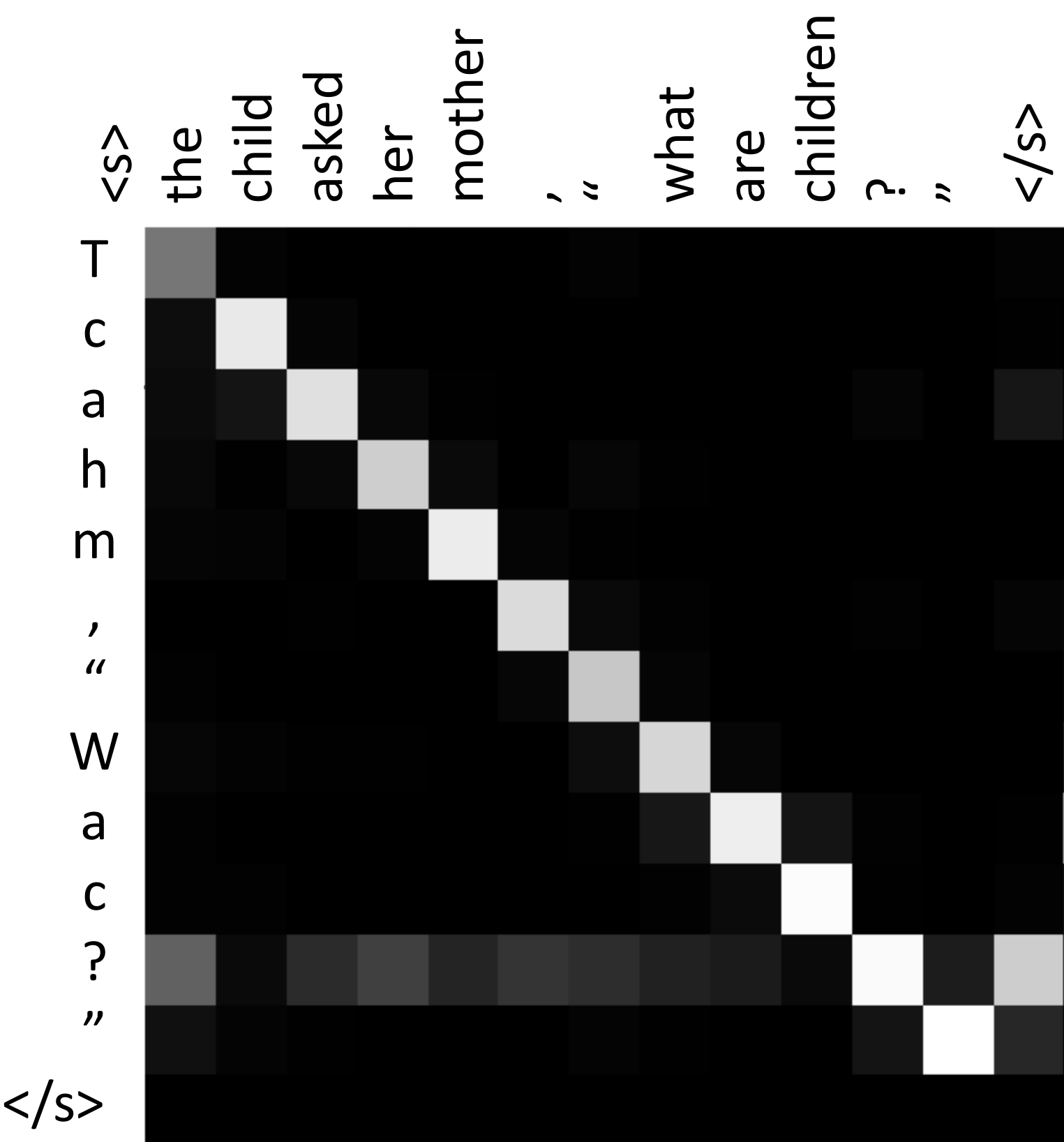}}
\subfloat[Case 6.]{\label{fig: Case 6.}
\includegraphics[width=0.23\textwidth]{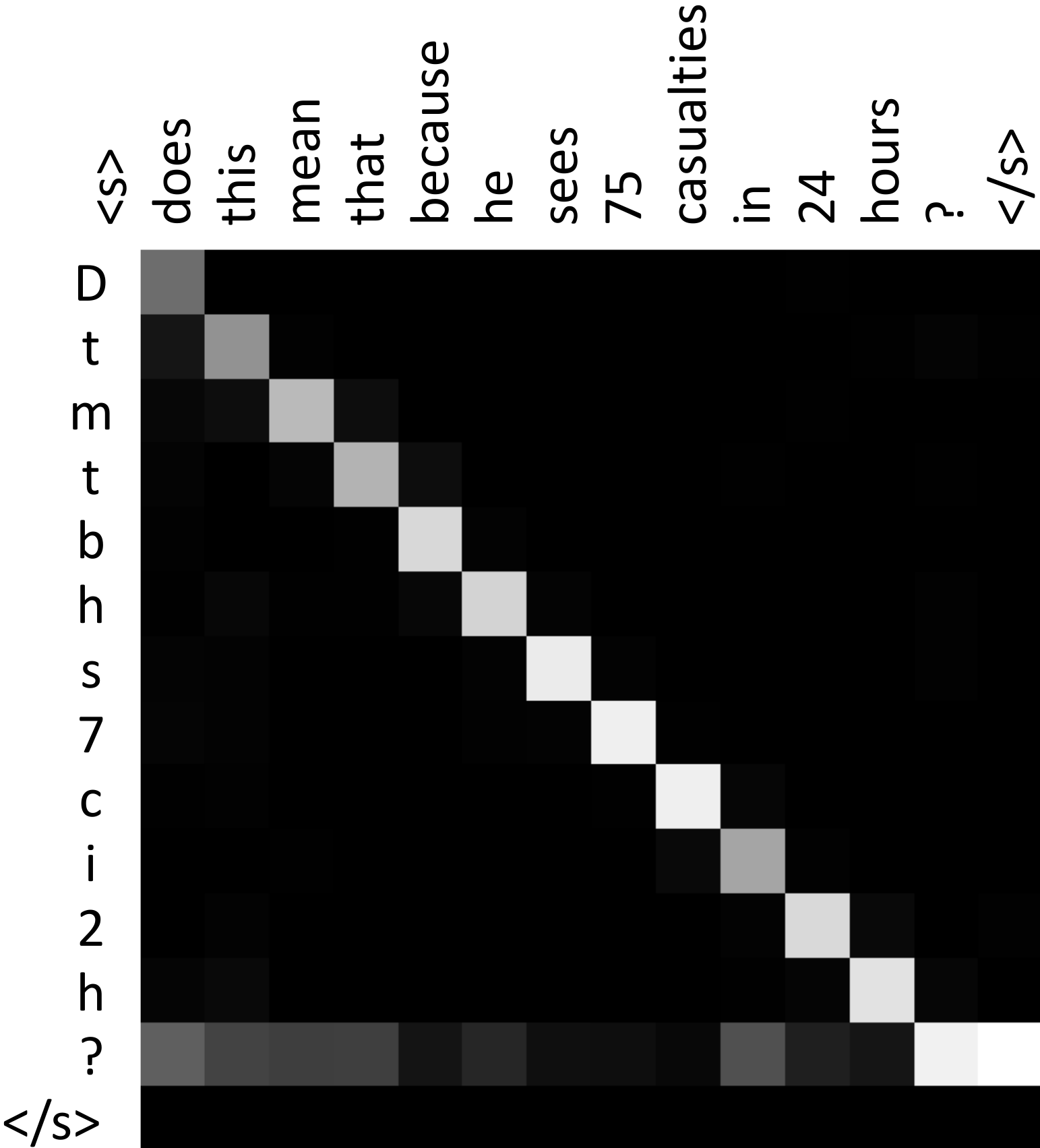}}
\caption{The heat map of the attention weights $\alpha_{t^\prime,t}$ in grayscale (0: black, 1: white)}
\label{figure: heat map.}
\end{figure}

Figure~\ref{figure: heat map.} visualizes the attention weights in the alignment between each pair of input password (y-axis) and generated mnemonic sentence (x-axis) for Case 5 and Case 6. 
Each pixel denotes the attention weight $\alpha_{t^\prime,t}$ of the $t$-th password character with regard to $t^\prime$-th target mnemonic word in grayscale (0: black, 1: white).    
The brighter the pixel is, the more important the password character is to the generation of the corresponding mnemonic word.
From the heat map, we can observe that DeepMnemonic concentrates on the important characters of an input password when it decodes individual target mnemonic words in the generation phase.
In addition, the attention layer also takes into account the contextual neighboring characters of each password for mnemonic sentence generation.
Thanks to the attention mechanism, DeepMnemonic automatically learns the alignment and discovers which characters in the input password are more important for generating a semantically meaningful mnemonic sentence.

\section{User Study}
\label{sec:User Study}

This section conducts a user study to analyze the usability of DeepMnemonic, and validates that the mnemonic sentences generated by DeepMnemonic are helpful for users to memorize passwords.

\begin{figure}[hbtp]
	\centering
	\includegraphics[scale=0.37]{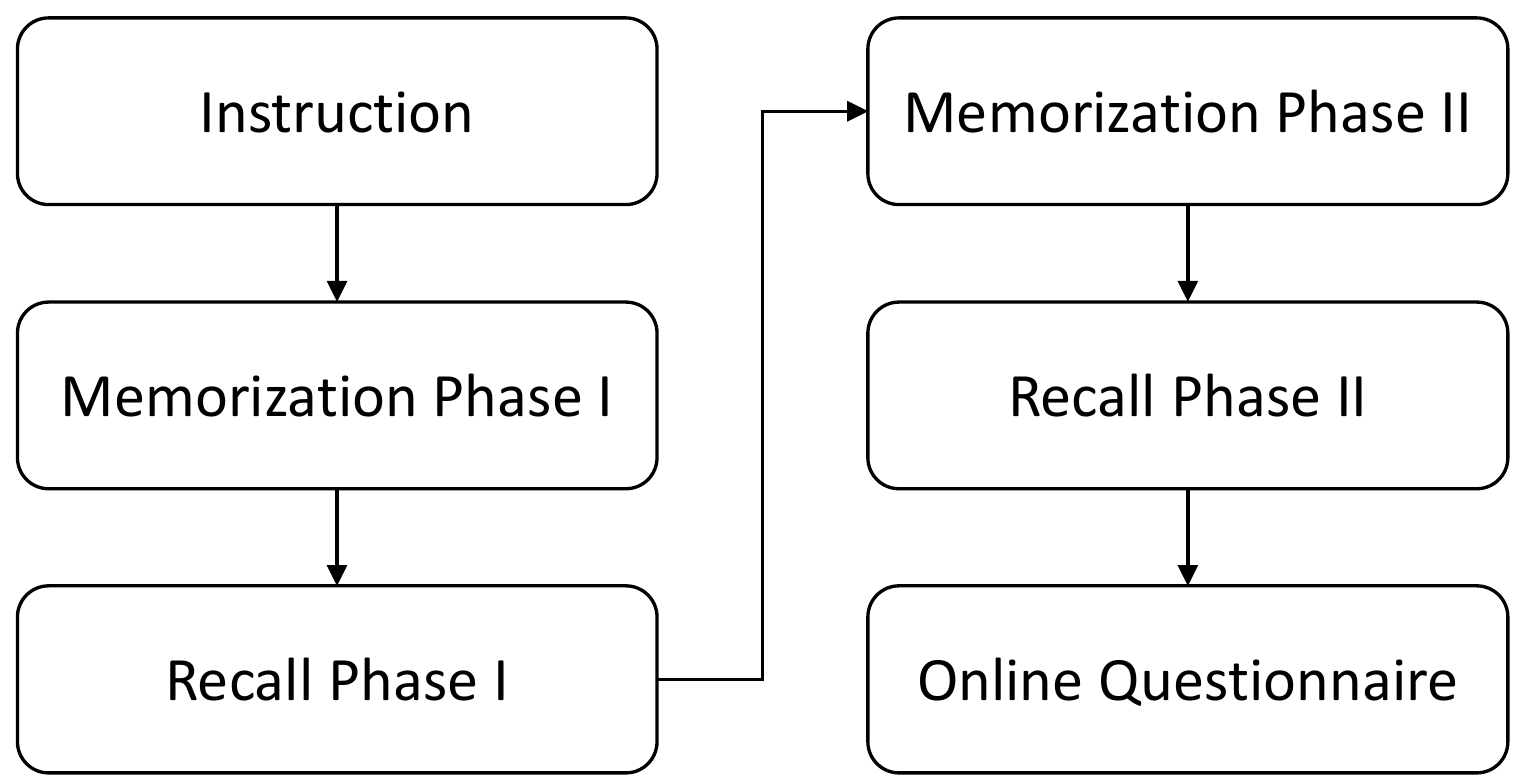}
	\caption{The user study procedure.}
	\label{fig: User study process overview.}
\end{figure}

\begin{figure*}
\centering
    \subfloat[Memorization time cost]{\label{fig: Memorization time cost.}
        \includegraphics[width=0.3\textwidth]{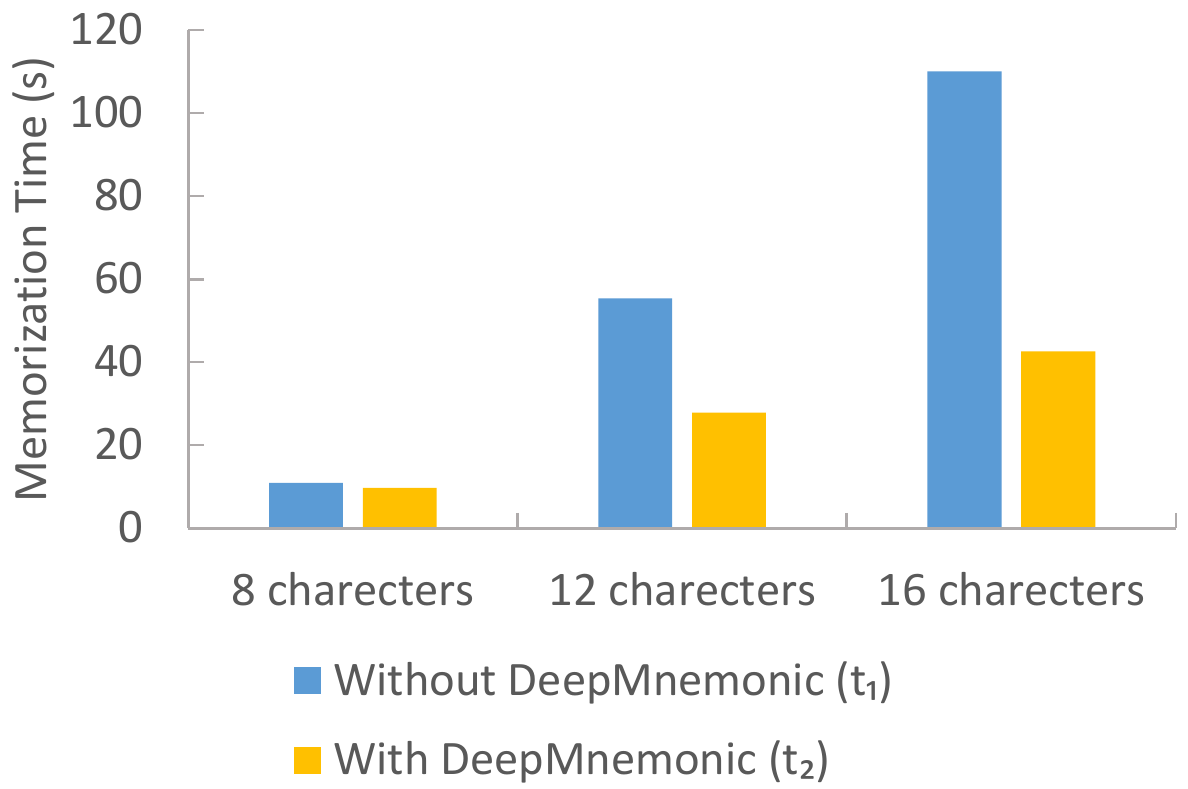}}
    \subfloat[Edit distance]{\label{fig: Edit distance.}
        \includegraphics[width=0.3\textwidth]{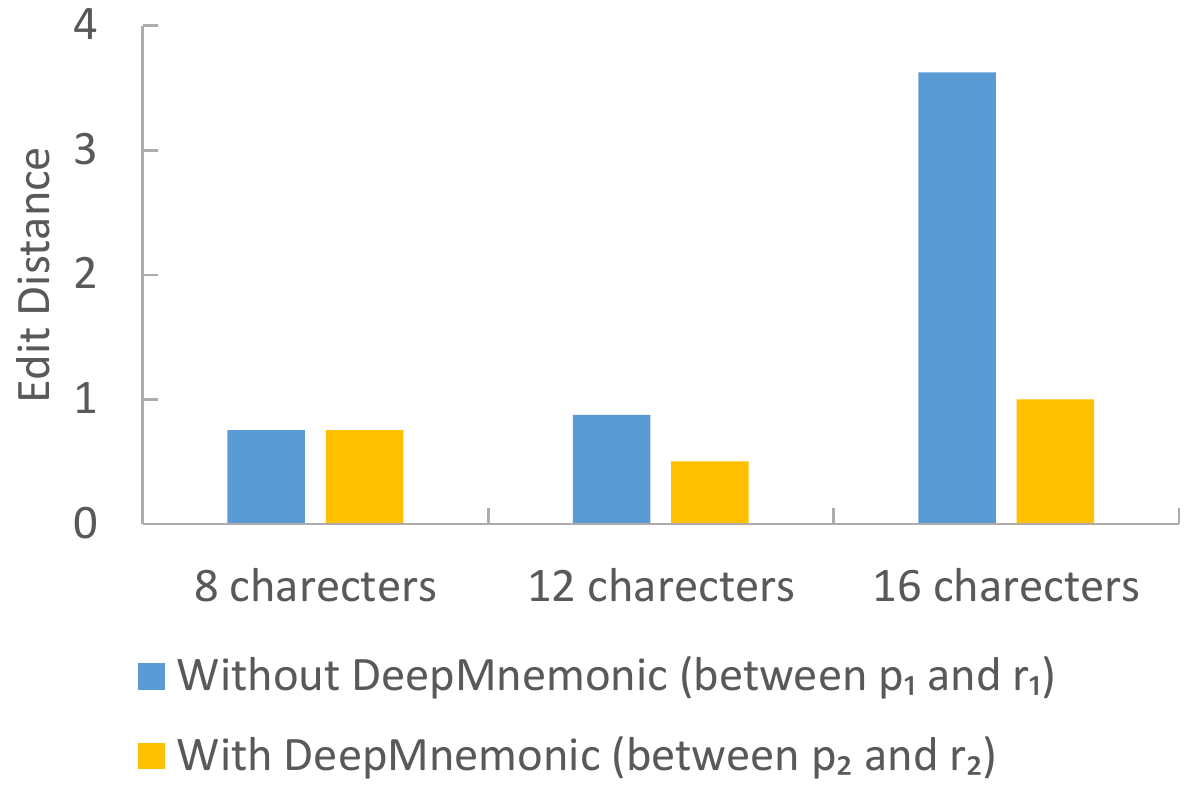}}
    \subfloat[Complete recall]{\label{fig: complete recall.}
        \includegraphics[width=0.3\textwidth]{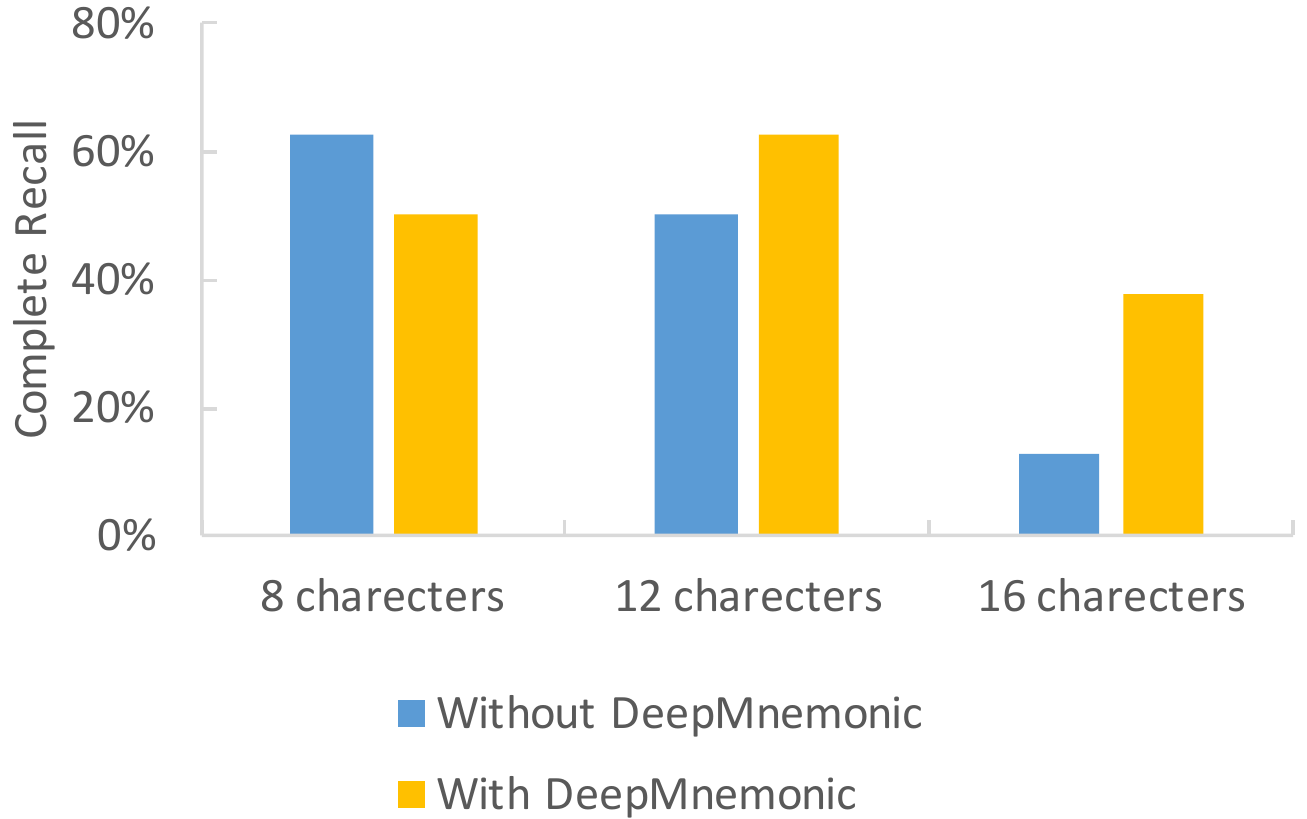}}    
    \caption{The comparison results of different user groups with and without using DeepMnemonic. We compare the memorization time cost (a), edit distance (b), and complete recall ratio (c) of the three groups without and with the aid of DeepMnemonic. For each comparison, the results for different password lengths are shown separately. }
    \label{figure: Comparison.}
\end{figure*}

\subsection{User Study Setting}

In this user study, we recruited 24 participants from universities and institutes, and randomly divided them into three groups, labeled as \textit{group 0}, \textit{group 1}, and \textit{group 2}.
The lengths of passwords assigned to the three groups were 8, 12, and 16, respectively.
This user study includes three main tasks, i.e., password memorization, password recall, and online questionnaire.
The password memorization and password recall tasks were conducted face-to-face in our lab, where we gave task instructions and measured the participants' performance using a timer.

Figure~\ref{fig: User study process overview.} illustrates the user study procedure.
In Memorization Phase I without the aid of mnemonic sentences, each participant was asked to memorize an assigned password $p_1$, where the time the participant used for memorization is $t_1$.
Then, after a period of time, in Recall Phase I, each participant was asked to recall and write down the assigned password $p_1$, and the recalled password is $r_1$.
In Memorization Phase II with the aid of the mnemonic sentences generated by DeepMnemonic, each participant was asked to memorize a different assigned password $p_2$, where the time the participant used for memorization is $t_2$.
Later in the Recall Phase II, the recalled version of $p_2$ is termed as $r_2$.
Finally, the online questionnaire was designed to evaluate the participants' experiences about the whole user study process.
Following the forgetting curve theory~\cite{ebbinghaus2013memory}, we set the time gap between password memorization and password recall to 48 hours\footnote{The detailed user study design as well as the ethical consideration can be found at \url{https://goo.gl/KtwuGB}}.

\subsection{User Study Results}

Figure~\ref{figure: Comparison.} shows the average results over all participants within each group (i.e., with the same password length) with and without the aid of the generated mnemonic sentences by DeepMnemonic.

Figure~\ref{fig: Memorization time cost.} shows the time costs in memorizing the passwords without ($t_1$) and with ($t_2$) the aid of mnemonic sentences.
Note that $t_2$ includes the time for memorizing the associated mnemonic sentence in addition to the password.
It is observed that $t_2$ is lower than $t_1$, especially for memorizing longer passwords. 
We conducted a statistical t-test which can work well with our small set of samples that approximate the normal distribution~\cite{dietterich1998approximate}\cite{bakirov2006student}.
The difference between the two sets of time costs is statistically significant at a significance level $\alpha=0.01$.
This indicates that, by using the mnemonic sentences generated by DeepMnemonic, the participants are able to memorize the passwords more quickly than without the aid of any mnemonics.
For \textit{group 2}, who memorized passwords of length 16, the observed difference is more significant, i.e., 110 seconds versus 42.6 seconds with $p_\text{value}=0.007$.
Clearly, DeepMnemonic shows its capability of assisting in memorizing passwords more effectively, and is especially helpful for memorizing relatively long passwords.

Figure~\ref{fig: Edit distance.} evaluates the password recall in terms of \textit{edit distance} between each pair of recalled password $r$ and assigned password $p$ (the smaller, the better).
Overall the average edit distance with the aid of DeepMnemonic is smaller than that without using DeepMnemonic.
The differences become even larger when the password lengths become longer.
For \textit{group 2} where the password length is 16, the average edit distance with the aid of DeepMnemonic is significantly smaller than that without the aid.
Our statistical test shows that the difference is significant given the level $\alpha = 0.01$. 
However, for shorter passwords in \textit{group 0} where password length is 8, the average edit distances with and without using DeepMnemonic are almost the same (0.75).
A possible explanation is that, when a given password is short, the efforts used to remember the generated mnemonic sentence and the password itself are comparable.
In such case, using mnemonic sentences may incur extra burdens for memorizing short passwords.

This can also be discovered from the complete recall ratio in Figure~\ref{fig: complete recall.} (the higher, the better). 
It shows that users can recall 8-character passwords better without the burden of memorizing extra mnemonic sentences. 
However, the helpfulness of mnemonic sentences becomes obvious for users to memorize 12-character and 16-character passwords.
Overall the mnemonic sentences generated by DeepMnemonic are effective in improving the performance of password recall, especially for longer/stronger passwords.

\begin{figure}[hbtp]
    \centering
    \includegraphics[scale=0.59]{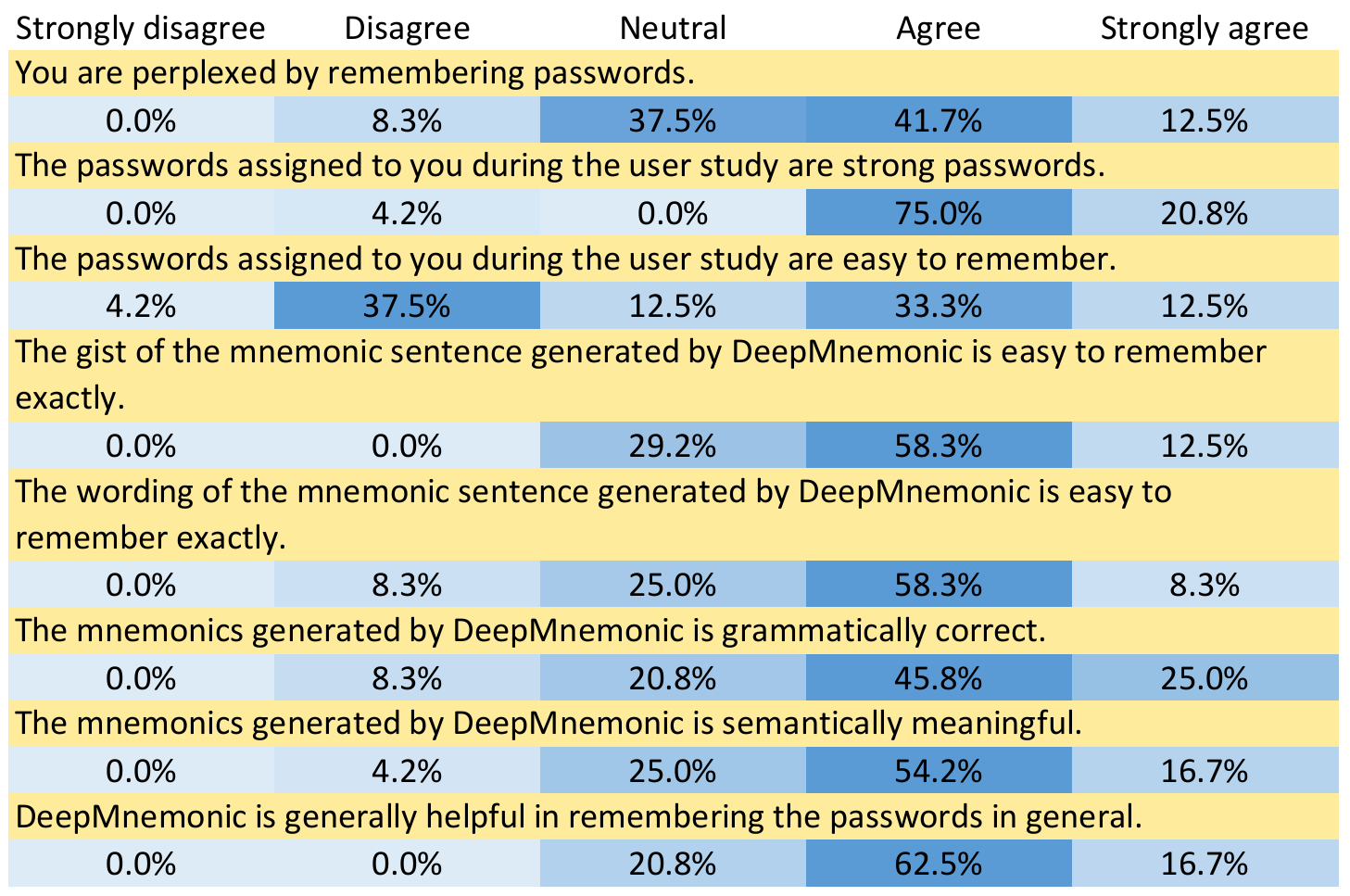}
    \caption{Online questionnaire analysis results.}
    \label{fig: Online Questionnaire results table.}
\end{figure}

In addition, we analyzed the results of online questionnaire to evaluate the user experience.
Figure~\ref{fig: Online Questionnaire results table.} shows that only 8.3\% of the participants indicated that they are \textit{not} perplexed by memorization of passwords, which clearly suggests that remembering passwords is indeed a difficult task.
Almost 96\% of participants agreed that the passwords assigned to them are secure and strong.
About 46\% of them considered their assigned passwords easy to remember.
We note that the majority of them come from \textit{group 0}, where their assigned passwords are not very long, and are thus not very difficult to remember.

A large portion of the participants agreed that both gist and exact words of each generated mnemonic sentence are helpful and easy to remember.
Almost 71\% and 67\% of participants agreed on the two points, i.e., ``gist is easy to remember'' and ``wording is easy to remember'', respectively.
With regard to grammatical correctness, about 8.3\% of participants noticed a few grammar errors existed in the generated sentences.
For example, in Case 6 of Table~\ref{table: Mnemonic Generation Examples}, the simple present tense in the sentence may not be rigorously reasonable.
The majority of the participants (70.9\%) agreed that the mnemonic sentences generated by DeepMnemonic are meaningful.
Overall, 79.2\% of participants recognized that DeepMnemonic is generally helpful for remembering the given passwords.

\section{Discussion}
\label{sec:Discussion}

In this section, we discuss several issues related to  the usability of DeepMnemonic.

\subsection{Learning from Multiple Strong Password Generation Rules}

In general, the successful application of DeepMnemonic to automatic mnemonic generation largely depends on the large-scale training data used for building the sequence-to-sequence learning model.
Recall that in our experiments, we generated pairwise password and mnemonic sentence training data from Webis-Simple-Sentences-17.
For each sentence in the dataset, we followed one strong password generation rule proposed in~\cite{kiesel2017large}, and thus concatenated all the first letter of each word and special characters in the sentence so as to create a password.
It is worth noting that DeepMnemonic is not limited to this single strong password generation rule adopted for constructing the pairwise training data.
It is possible that DeepMnemonic can also be trained on other datasets constructed using different password generation rules, such as concatenating the last letter of each word instead of the first letters in deriving passwords.
Thanks to the use of the \textit{sequence-to-sequence} learning model, DeepMnemonic is able to learn language generation patterns automatically from the training data, and map passwords to mnemonic sentences in testing.
Therefore, it is possible to train multiple encoder-decoder language generation models in DeepMnemonic using different password generation rules.
As a result, users have flexibility to choose from multiple generated mnemonic sentences for a given password, which may further improve the usability of DeepMnemonic.

\subsection{Baseline Selection}
The $n$-gram language model used in our comparison evaluation is one of the most well-known language generation models.
We have tested various values of $n$ ($n\in[1, 4]$).
Since the bigram language model achieved the best performance among different $n$, it was selected as the baseline.

We discovered that when using a unigram language model to map from a password sequence to a mnemonic sequence, the model cannot leverage on the context, i.e., preceding words. 
As a result, the words that appear more frequently in training data are considered more important, and are then assigned with a higher probability for prediction.
For example, given a password character ``t'', it is very likely that the unigram language model would generate the word ``the'', regardless of the preceding contextual words.
In addition, trigram ($n=3$) and 4-gram based language models built on the pairwise training data often suffer from sparse co-occurrence information, and output the unknown token $<${\scriptsize UNK}$>$.
It is more likely that these models, compared to the bigram models, generate empty list of candidate words from two or more contextual preceding words and a target password character.
According to our experimental results, when $n$ is larger than 2, the language models have more difficulty in generating complete sentences.

\subsection{Evaluation Metric}

BLEU is one of the most popular automated metrics for evaluating the \textit{quality} of machine translation~\cite{papineni2002bleu} in natural language processing.
The quality is often considered to be the matching degree between each pair of generated sentence and ground-truth sentence.
The closer the generated sentence is to the ground-truth, the better the language generation is.
Our quantitative experimental results show that DeepMnemonic significantly outperforms the well-established bigram language model in terms of BLEU.

We note that BLEU is a good metric to evaluate the quality of machine translation, where each generated or translated sentence should be close to the reference translation (ground truth) as much as possible.
The state-of-the-art BLEU score is 34.8 for machine translation using deep neural networks~\cite{sutskever2014sequence}. 
However, BLEU may not be a \textit{fair} metric for evaluating the password mnemonic generation, which aims to assist users in password memorization.
It is true that the generated mnemonic sentences sometimes do not match perfectly to the reference sentences (ground truth), which may lead to poor BLEU scores.
However, it is possible that such mnemonic sentences are still helpful for memorizing passwords, as long as they are grammatically correct and semantically meaningful by themselves, as shown in Table~\ref{table: Mnemonic Generation Examples}.
The metric MP proposed in Section~\ref{MP} is thus designed to compliment BLEU in terms of mnemonic generation evaluation.

\subsection{User Study Scale and Population}

In our user study, we recruited 24 participants to evaluate the usability of DeepMnemonic. 
Despite the small scale, the user study shows that DeepMnemonic is helpful in assisting users in better memorizing the passwords, especially when passwords are relatively long. 
A limitation of this user study lies in the age distribution of the participants. 
The ages of the participants range from 24 to 40, which are not representative of younger or older population.
Indeed, this is also a challenge in many other user studies~\cite{li2017you}.
Our user study suggests that DeepMnemonic is generally helpful for young and middle-aged people.
It remains a future work to evaluate the DeepMnemonic against other age groups.

\subsection{Lowercase and Uppercase Password Characters}
DeepMnemonic does not handle lowercase and uppercase characters when generating mnemonic sentences, and thus cannot help users distinguish between them. 
We have tried to build a variant DeepMnemonic model, which can generate sentences that differentiate lowercase and uppercase characters. 
However, its performance (e.g., BLEU and MP) is not as good as that of current DeepMnemonic model. 
We plan to deal with this lowercase and uppercase character issue in the future work.   
In addition, users may have multiple passwords across various platforms.
It would be interesting to extend DeepMnemonic to help users memorize their passwords across different platforms. 

\subsection{Generating Mnemonics in Different Languages}

Currently, the proposed DeepMnemonic has been shown to be effective in generating mnemonics for passwords in English.
Similar to the \textit{sequence-to-sequence} machine translation task in natural language processing~\cite{bahdanau2014neural}, the underlying encoder-decoder model of DeepMnemonic can encode any input password into a semantic vectorial representation and decode the representation into a semantically meaningful mnemonic sentence in the target language.
When suitable training datasets in other languages (e.g., Chinese) are available, DeepMnemonic can be easily adapted to the mnemonic sentence generation in those language.

\section{Related Work}
\label{sec:related work}

\subsection{Strong Password Evaluation and Generation}

A lot of efforts have been made in the past to assess the strength of passwords or to generate strong passwords. 
The strength of a password can be typically evaluated by two common metrics, namely, entropy and guessability. 
The entropy measures how unpredictable a password is by considering the length of password and the distribution of characters in the password.
One limitation of the entropy-based measurement~\cite{nist} is that it only supplies users with rough approximations of password strength~\cite{weir2010testing}\cite{kelley2012guess}\cite{ur2012does}.
In contrast, guessability-based measurement, which is defined as the number of guesses required to break a password, has become increasingly popular.  
To evaluate the guessability of a password, one key step is to identify an algorithm for password cracking, such as the Probabilistic Context-Free Grammar model (PCFG)~\cite{weir2009password}\cite{veras2014semantic}\cite{wang2016fuzzypsm} and the Markov $n$-grams model~\cite{castelluccia2012adaptive}\cite{ma2014study}. 
These cracking algorithms exploit the password distributions that are derived from various password datasets disclosed in previous security incidents. 
It is revealed that different password datasets collected from different user groups demonstrate different distributions~\cite{wang2019birthday}\cite{wang2017understanding}.
Then, the guessability of the password can be measured using the cracking algorithm.

Previous studies have evaluated the password strength by measuring the popularity of passwords. 
Schechter et al.~\cite{schechter2010popularity} evaluated user-chosen passwords by identifying undesirably popular passwords. 
The passwords within certain popularity threshold are considered secure under probabilistic attacks. 
It is argued that the existing password creation policies can be replaced with popularity limits so as to strengthen both security and usability for user authentication systems.

Strong password generation aims to strengthen passwords via changing or adding characters to the passwords.
Generation of persuasive text passwords is one of such approaches, which inserts one to four characters at random positions in a given password~\cite{forget2008improving}\cite{forget2008persuasion}.
Recently, Houshmand and Aggarwal~\cite{houshmand2012building} presented an \textit{analyze-modify} method to generate strong passwords.
They first evaluated password strength using PCFG.
For identified weak passwords, they modified them based on a set of editing rules.
One limitation of strong password generation is that it does not take into consideration the usability or memorability of passwords.
As a consequence, the generated passwords may not be user-friendly in practice.

\subsection{Memorable Password Generation}

A variety of strategies have been employed to create easy-to-remember passwords by service providers.
One strategy is to generate pronounceable passwords~\cite{ppg}\cite{npg}. 
For example, pronounceable password ``kilakefe52'' is comprised of a random sequence of vowel-consonant pairs~\cite{ppg}.
Another strategy is to create memorable passwords that comprise multiple components with certain meanings~\cite{mpg}.
For example, password ``\#FreDDi17\%'' can be divided into three meaningful parts, i.e., ``\#'' (identity), ``FreDDi'' (name), and ``17\%'' (percentage).
The third strategy aims to generate passwords by simply concatenating a list of random words via hyphen character~\cite{npg2}. 
The benefit of this strategy is twofold: 
(i) It is usually difficult to guess unrelated words of the generated passwords; 
and (ii) Longer passwords often lead to higher entropy and security~\cite{Shay:2012:CHB:2335356.2335366}.
Although the generated passwords via the aforementioned strategies are memorable, their strengths have not been systematically evaluated yet,  and some of them are vulnerable to password attacks~\cite{yang2016empirical}.

One typical approach to generate memorable yet strong passwords is to rely on meaningful language expressions, making use of specific parts of given reference sentences or phrases according to certain generation rules~\cite{Kuo:2006:HSM:1143120.1143129}\cite{yang2016empirical}\cite{kiesel2017large}.
For example, password ``\textit{Ilwm.}'' can be generated by joining the first character of each word and the punctuation in the following sentence: ``\textit{I love watching movies.}''
The expression-based passwords generated by this approach are often supposed to be stronger than those selected intuitively by users~\cite{yan2004password};
given reference sentences or phrases, the usability cost for memorizing the generated passwords is almost comparable to memorizing those intuitive ones.
Yang et al.~\cite{yang2016empirical} demonstrated that the security level of an expression-based password was largely affected by its generation rules, which was also validated by Kiesel et al.~\cite{kiesel2017large}.
However, it is not clear how various generation rules at different security levels affect the usability costs for memorizing the generated passwords.
Due to users' behavioral tendency toward picking up easy-to-remember reference sentences~\cite{forget2008improving}, the generated memorable passwords using such human-selected language expressions may be easy-to-guess by attackers.

Researchers have studied various tips for creating mnemonic passwords, including sentence substitution (SenSub), key-board change (KbCg), using a formula (UsForm), and special character insertion (SpIns)~\cite{ye2019empirical}. 
Alphapwd~\cite{song2019alphapwd} is a memorable password generation strategy based on password shapes. 
In order to create memorable passwords, Alphapwd requires user to remember a shorter sequence of letters that are shown in larger size on top of a normal keyboard; a user may derive his/her strong password easily from the keyboard following the strokes of the shorter sequence of large-size letters.
The strengths of these generated passwords are evaluated using probabilistic attack algorithms such as the PCFG algorithm and MarKov model trained on previous disclosed password datasets. 
Ghazvininejad and Knight~\cite{ghazvininejad2015memorize} proposed to generate passwords in the form of English sequences, called \textit{passphrase}, whose lengths range from 31.2 to 87.7 characters on average. 
However, experiments revealed that it is difficult for common users to reproduce the exact wording of passphrases, even if they manage to remember the gist of these sentences~\cite{yajam2016papiapass}. 
Being complementary to these memorable password generation methods, DeepMnemonic is designed to generate mnemonic sentences for any given passwords instead of generating new passwords.

\subsection{Password Mnemonics}

To improve the memorability of given strong passwords, various types of hints or mnemonic tools have been used.
Atallah et al.~\cite{atallah2001natural} is the first to propose the concept of using funny jingles for memorizing a randomly generated password.
One challenge of this proposal is the generation of related funny jingles.
Several other efforts were made to assist users in remembering passwords via external tools, such as helper card~\cite{topkara2007passwords}, hint image~\cite{fukumitsu2010proposal}\cite{juang2012using}, and engaging game~\cite{doolani2016improving}.
Jeyaraman and Topkara ~\cite{jeyaraman2005have} proposed to match given passwords to textual headlines or their variant versions selected from a given corpus to assist users in remembering passwords.
One drawback of this approach is that the variant headlines generated may be syntactically incorrect or semantically inconsistent.
Due to the limited length of the headline text, it can only generate hints for short passwords and largely fails for long and strong passwords.
The memorability of the passwords generated by this approach has not been evaluated.

\subsection{Natural Language Generation}
In natural language processing, statistical language models are used to generate meaningful sentences by computing joint probabilities of sequences of words from a dictionary of a given language.
One of the dominant methods for probabilistic language modeling is the $n$-gram language model~\cite{manning1999foundations}, which is a non-parametric learning algorithm.
It relies on the preceding sequence of $n-1$ words to estimate the conditional probability for predicting (generating) the current $n$-th word.

Recently, neural network based language models have become increasingly popular in natural language generation tasks.
Bengio et al.~\cite{bengio2003neural} proposed a generic neural probabilistic language model, which can simultaneously learn the distributed representation of each word and joint probability function of sequences.
Sutskever et al.~\cite{sutskever2014sequence} proposed a sequence-to-sequence neural language model to address the machine translation problem.
One key benefit of their model is that it can automatically generate a translated sentence in the target language, given a sentence in the source language. 
In order to improve machine translation, Bahdanau et al.~\cite{bahdanau2014neural} introduced an attentive alignment strategy to enhance the sequence-to-sequence neural language model, which learned to dynamically pay more attention to salient parts of the input sentences when generating the translated sentences.

In this paper, we exploit natural language translation techniques to generate human-readable and semantically meaningful mnemonic sentences from any given passwords so as to help users memorize strong passwords.

\section{Conclusion}
\label{sec:Conclusion}

In this work, we have proposed DeepMnemonic, a deep neural network based approach to automatic generation of mnemonic sentences for any given textual passwords.
DeepMnemonic builds upon an attentive encoder-decoder language generation framework, and works by \textit{translating} an input sequence of password characters to a natural language sentence of mnemonic words.
DeepMnemonic is designed to bridge the gap between the strong password generation and the usability of strong passwords.
Experimental results show that DeepMnemonic is capable of generating semantically meaningful mnemonic sentences.
A user study is conducted to evaluate the usability of DeepMnemonic, which shows that the generated mnemonic sentences are helpful in memorizing strong passwords.
Specifically, with the aid of DeepMnemonic, the time used for remembering a password is largely reduced, and the password recall quality is also significantly improved.
In the future, we plan to train DeepMnemonic using more comprehensive and diverse training data.


\bibliographystyle{IEEEtran}
\bibliography{TDSC-2019-06-0296}


\begin{IEEEbiography}[{\includegraphics[height=1.25in]{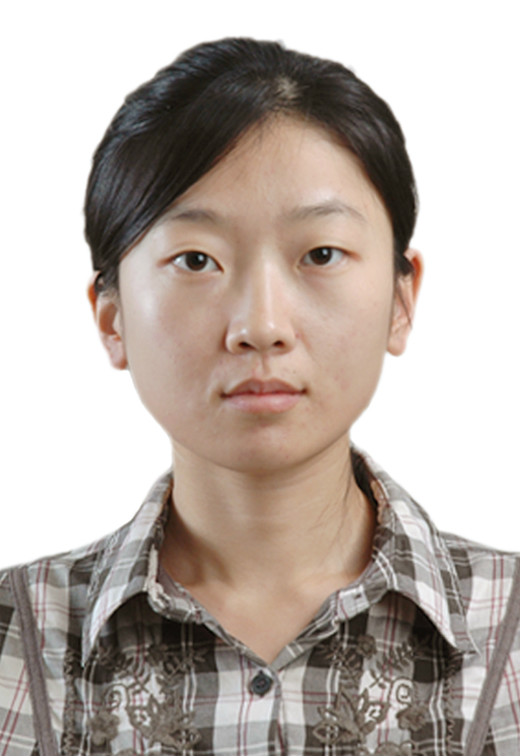}}]{Yao Cheng}
is currently a senior researcher at Huawei International in Singapore.
She received her Ph.D. degree in Computer Science and Technology from University of Chinese Academy of Sciences. 
Her research interests include security and privacy in deep learning systems, blockchain technology applications, Android framework vulnerability analysis, mobile application security analysis, and mobile malware detection.
\end{IEEEbiography}

\begin{IEEEbiography}[{\includegraphics[height=1.25in]{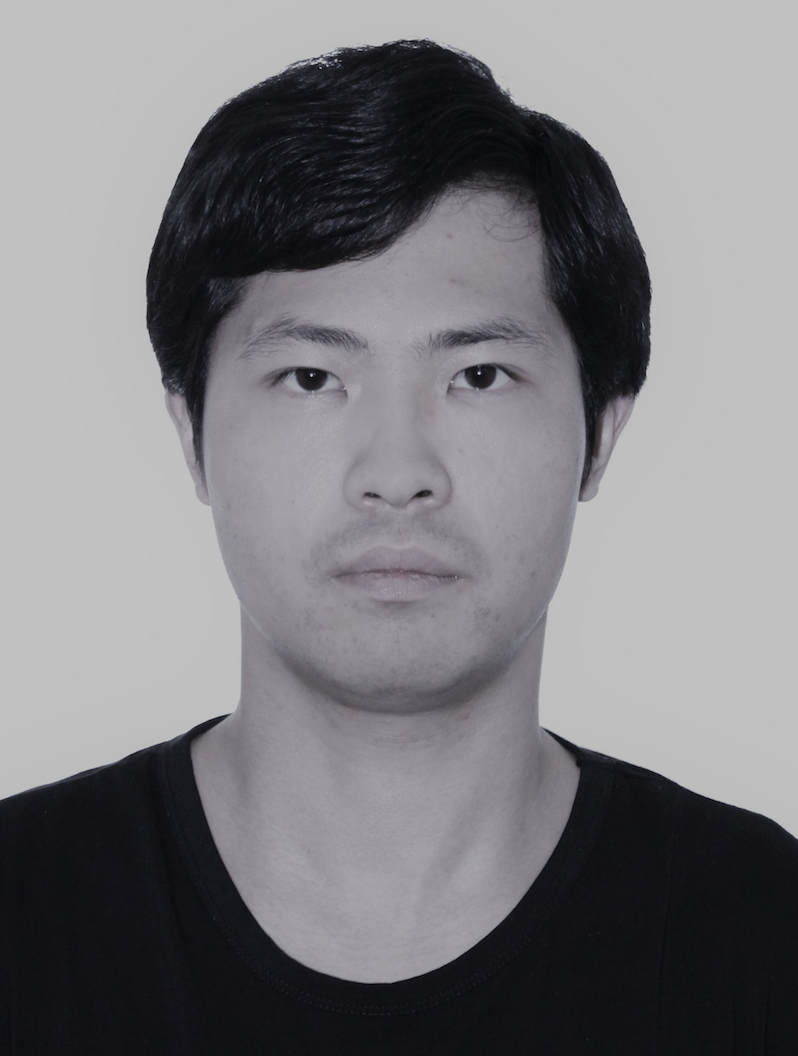}}]{Chang Xu}
is currently a Postdoctoral Fellow at Data61, CSIRO, Australia. He received his PhD degree in Computer Science from Nanyang Technological University in March 2017. His current interests include robust and explainable deep neural models for natural language processing.
\end{IEEEbiography}

\begin{IEEEbiography}[{\includegraphics[height=1.25in]{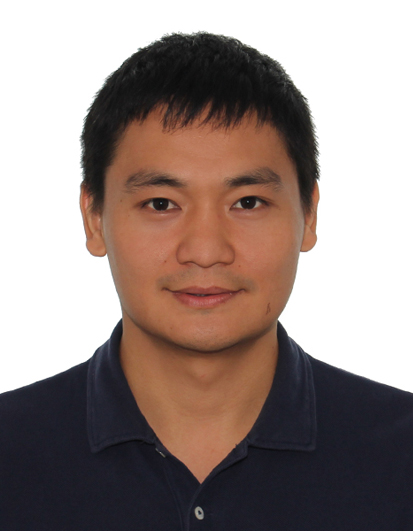}}]{Zhen Hai}
received the PhD degree in Computer Science and Engineering from Nanyang Technological University  in 2014.
He has been with the Institute for Infocomm Research, A*STAR, Singapore since 2015.
His research interests include natural language processing, text mining, sentiment analysis, information security, and machine learning.
He has been invited to serve on the program committees of leading conferences including SIGIR, ACL, EMNLP, AACL, AAAI, IJCAI, CIKM, etc.
\end{IEEEbiography}

\begin{IEEEbiography}[{\includegraphics[height=1.25in]{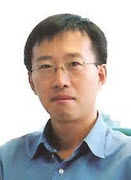}}]{Yingjiu Li}
is currently a Ripple Professor in the Department of Computer and Information Science at the University of Oregon. 
His research interests include IoT Security and Privacy, Mobile and System Security, Applied Cryptography and Cloud Security, and Data Application Security and Privacy. 
He has published over 140 technical papers in international conferences and journals, and served in the program committees for over 80 international conferences and workshops, including top-tier cybersecurity conferences. 
\end{IEEEbiography}
\end{document}